\documentclass{aa}
\usepackage[varg]{txfonts}
\usepackage{graphicx}
\usepackage{natbib}
\usepackage{textcomp} 
\usepackage[normalem]{ulem}

\usepackage[bookmarks, breaklinks=true, colorlinks=true,
  citecolor=blue,urlcolor=blue,filecolor=blue, linkcolor=blue,
  pdftitle={Evolution of critically-rotating accretors in Algols},
  pdfauthor={R. Deschamps}, pdfkeywords={Binaries: general, Radiative
    transfer, Stars: circumstellar matter, Methods:
    numerical}]{hyperref}

\bibpunct{(}{)}{;}{a}{}{,}

\begin{document}

\title{Non-conservative evolution in Algols: where is the matter?}

\author{R. Deschamps\inst{1,2} \and K. Braun\inst{2} \and
  A. Jorissen\inst{2} \and L. Siess\inst{2} \and M. Baes\inst{3} \and
  P. Camps\inst{3}}

\offprints{Romain.Deschamps@ulb.ac.be}

\institute{ European Southern Observatory, Alonso de Cordova 3107,
  Casilla 19001, Santiago, Chile \and Institut d'Astronomie et
  d'Astrophysique, Universit\'e Libre de Bruxelles, ULB, CP 226,
  B-1050 Brussels, Belgium \and Sterrenkundig Observatorium,
  Universiteit Gent, Krijgslaan 281 S9, B-9000 Ghent, Belgium}

\abstract {There is gathering indirect evidence suggesting
  non-conservative evolutions in Algols. However, the systemic
  mass-loss rate is poorly constrained by observations and generally
  set as a free parameter in binary-star evolution
  simulations. Moreover, systemic mass loss may lead to observational
  signatures that are still to be found.} 
{Within the `hotspot' ejection mechanism, some of the material that is
  initially transferred from the companion star via an accretion
  stream is expelled from the system due to the radiative energy
  released on the gainer's surface by the impacting material. The
  objective of this paper is to retrieve observable quantities from
  this process, and to compare them with observations.} 
{We investigate the impact of the outflowing gas and the possible
  presence of dust grains on the spectral energy distribution
  (SED). We used the 1D plasma code \textsc{Cloudy} and compared the
  results with the 3D Monte-Carlo radiative transfer code
  \textsc{Skirt} for dusty simulations. The circumbinary
  mass-distribution and binary parameters are computed with
  state-of-the-art binary calculations done with the \textsc{Binstar}
  evolution code.}
{The outflowing material reduces the continuum flux-level of the
  stellar SED in the optical and UV. Due to the time-dependence of
  this effect, it may help to distinguish between different ejection
  mechanisms. Dust, if present, leads to observable infrared excesses
  even with low dust-to-gas ratios and traces the cold material at
  large distances from the star. By searching for such dust emission
  in the WISE catalogue, we found a small number of Algols
  showing infrared excesses, among which the two rather surprising
  objects SX~Aur and CZ~Vel. We find that some binary B[e] stars
    show the same strong Balmer continuum as we predict with our 
    models. However, direct evidence of systemic mass loss is probably
    not observable in genuine Algols, since these systems no longer
    eject mass through the hotspot mechanism. Furthermore, owing to
    its high velocity, the outflowing material dissipates in a few
    hundred years.  If hot enough, the hotspot may produce highly
  ionised species such as \ion{Si}{iv} and observable characteristics
  that are typical of W~Ser systems. }
{Systemic mass loss, if present, leads to clear observational
  imprints. These signatures are not to be found in genuine Algols
  but in the closely related $\beta$~Lyr\ae{}s, W~Serpentis
  stars, Double Periodic Variables, Symbiotic Algols and binary B[e]
  stars. We emphasise the need for further observations of such
  objects where systemic mass loss is most likely to
  occur.}


\maketitle

\section{Introduction}
\label{sec:introduction}

Algol systems are short-period, semi-detached, binary systems (with
orbital periods from several hours to tens of days) generally composed
of a hot (B-A) main-sequence star and a less massive but more evolved
companion overfilling its Roche lobe and transferring mass via Roche
lobe overflow \citep[RLOF;][]{1955AnAp...18..379K,
  1983ApJS...52...35G}.

Although Algol systems have been studied since the 18$^{\mathrm{th}}$
century \citep{1783RSPT...73..474G} and seem to be rather simple
objects within the zoo of binary systems, it is still unclear whether
the mass transfer can be fully conservative, or if systemic mass loss
is a universal feature of Algols as it has been inferred already by,
e.\,g., \citet{1955ApJ...121...71C}. It has been shown that systemic
angular-momentum losses (and thus mass losses) must be involved to
reproduce the current state of some observed Algols. For example,
\citet{1974A&A....36..113R} found that only a non-conservative
evolution can reproduce the system
AS~Eridani. \citet{1975MmSAI..46..217M} noticed that Algol models must
loose at least 40\% of their mass in order to reproduce observed
properties. Eventually, \cite{1993MNRAS.262..534S} concluded
that the Algol prototype $\beta$~Per must have lost 15\% of its
initial total mass during its evolution and 30\% of its total angular
momentum. Statistically, \cite{1979Ap&SS..64..177C} found that
semi-detached binaries have lower orbital angular momentum than
detached systems of the same total mass. Finally, many other clues
(e.\,g., comparison of grid of models with observations or
population synthesis analysis; \citealt{1980A&A....83..217M,
  2001ApJ...552..664N, 2011A&A...528A..16V}) converge in favour of
non-conservative evolution.

Different mechanisms have been proposed to explain systemic mass
loss. They may be classified into four groups: bipolar jets
\citep{2000A&A...358..229U, 2002A&A...391..609U, 2007A&A...463..233A,
  2011BSRSL..80..689L}, enhanced winds \citep[chapter~4
  of][]{2009pfer.book.....M,1992MNRAS.256..269T}, losses through the
outer Lagrangian point $\mathcal{L}_{3}$ \citep{2007ARep...51..836S},
and `hotspot'. The latter mechanism appears to be the most promising,
because the mass-loss rates derived from the first three mechanisms
are too low to account for the inferred systemic mass loss of some
Algols. In this scenario, the mass loss is driven by the radiation
pressure of a hotspot \citep[][hereafter Paper~I]{2008A&A...487.1129V,
  2010A&A...510A..13V, 2013A&A...557A..40D} that forms when the stream
of material coming from the donor via RLOF hits the accreting star and
releases energy to heat the stellar surface and increase the
luminosity locally. This hotspot is observed in some Algol-like
stars (e.\,g., \citealt{1990Ap&SS.173...77B, 1995ApJ...447..401W,
  1999ApJS..123..537R}, in case of direct accretion on the stellar
surface and \citealt{2002AN....323...87H, 2007MNRAS.379.1533S} in case
of accretion on the edge of a disc).

In the line of \cite{2008A&A...487.1129V}, \cite{2013A&A...557A..40D}
re-derived a model for systemic mass loss via hotspot based on energy
conservation at the impact location and showed that for a typical
initial 3.6~M$_{\odot}$ + 6~M$_{\odot}$ system with an initial orbital
period P~=~2\fd5, a large amount of matter (of the order of
1.4~M$_{\odot}$, i.\,e.\ 15~\% of the initial total mass of the
system) can be lost during the short-lasting ($2 \times 10^{5}$ years)
rapid ($\dot{M}_{\mathrm{RLOF}} = 1 \times 10^{-5} - 1 \times
10^{-4}$~M$_{\odot}$ yr$^{-1}$) mass-transfer phase. This total amount
of mass lost is in good agreement with the value inferred by
\cite{1993MNRAS.262..534S} to reconcile Algol models with
observations. Using the mass-loss history, and assuming the wind to
follow the prescription of \cite{1975ApJ...195..157C} for line-driven
winds, a density profile for the material surrounding the system was
derived.

In this paper, we confront this prediction of systemic mass loss to
observations by converting the density profile into observable
quantities. We will thus perform an analysis of light emission and
absorption by this material using the codes \textsc{Cloudy}
\citep{2013RMxAA..49..137F} and \textsc{Skirt}
\citep{2003MNRAS.343.1081B, 2011ApJS..196...22B,
  2015A&C.....9...20C}. Section~\ref{sec:model} briefly recaps and
discusses the assumptions made to obtain the density profile and
presents the characteristics of the model system and the set-up of the
two codes. The results are presented in Sect.~\ref{sec:results}. We
then confront our findings to Algol observations and broaden our study
to promising Algol-related objects in Sect.~\ref{sec:obs} before
concluding in Sect.~\ref{sec:conclusion}.

\section{The model system}
\label{sec:model}

\subsection{Density profile of the out-flowing matter}
\label{sec:rho}

\begin{figure}[t]
  \centering\includegraphics[width=0.48\textwidth]{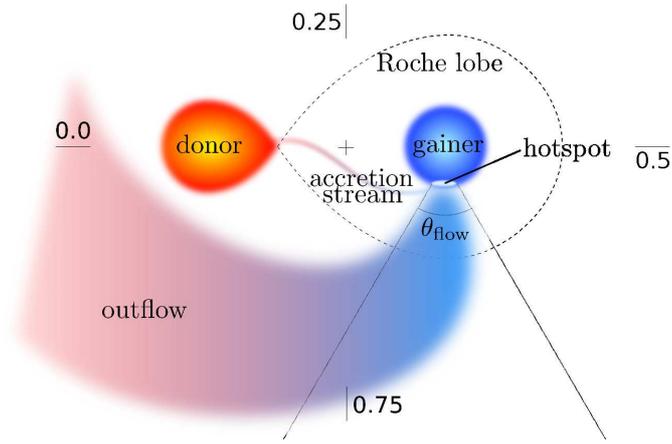}
  \caption{Schematic overview of an Algol system undergoing systemic
    mass loss due to a hotspot on the surface of the gainer. The ticks
    and numbers indicate the different phases (0.0: primary eclipse).}
  \label{fig:global}
\end{figure}

We follow the ballistic trajectory (RLOF-stream) and thus the specific
kinetic energy of the material escaping from the donor star via
Roche-lobe overflow through the Lagrangian point $\mathcal{L}_{1}$ and
accreting onto the gainer star by direct impact (Paper~I).  When
impacting onto the stellar surface, a fraction of this energy heats up
the perturbed stellar surface, forming a hotspot (see
Fig.~\ref{fig:global}), and the remainder is radiated away. Mass loss
is assumed to occur whenever the total luminosity at the impact
location (stellar + hotspot luminosity) exceeds the local Eddington
luminosity. The latter is computed using the gainer upper layers'
opacity.

We assume the wind (hereafter outflow) emitted by the hotspot to be
line-driven, and follow the \citet{1975ApJ...195..157C} prescription
to compute the outflow speed $v_{\mathrm{flow}}$. Due to the orbital
motion, the outflow will exhibit a spiral shape
(Fig.~\ref{fig:global}; see Sect.~\ref{sec:spiral}). The acceleration
of the particles moving in this spiral outflow will be considerably
reduced once the matter leaves the ``irradiation'' cone defined by the
opening angle $\theta_{\mathrm{flow}}$ (whose estimate is provided in
Appendix~\ref{ap:opening_angle}). It is important to stress that a
mass element ejected from the hotspot will escape the system, even if
the outflowing material is not constantly lit up by the hotspot,
because it exceeds its escape velocity after a small fraction of the
orbital period\footnote{As presented in Fig~\ref{fig:density}, the
  escape velocity is reached at a fixed distance from the star and
  therefore after a fixed amount of time.} ($0.038 \times
P_{\mathrm{orb}}$ at the beginning of the mass-transfer phase, $0.011
\times P_{\mathrm{orb}}$ at the end of the non-conservative phase),
during which the change of the orientation of the hotspot is
negligible.

The density of the material surrounding the system at a given time $t$
and distance $r$ is calculated using the continuity equation (see
Paper~I for a complete description):
\begin{equation}
  \label{eq:rho}
  \rho(r,t) = \frac{\dot{M}_{\mathrm{flow}}(r,t)}{4\pi f_\mathrm{c} r^{2}
    v_{\mathrm{flow}}(r,t)} , 
\end{equation}
where $\dot{M}_{\mathrm{flow}}$ is the systemic mass-loss rate
calculated with the \textsc{Binstar} code
(\citealt{2013A&A...550A.100S}, Paper~I) and $4\pi f_\mathrm{c}$ is
the solid angle into which the material is ejected. $f_\mathrm{c}$ is
the covering factor defined by Eq.~\ref{eq:sigma} and is related to
the opening angle $\theta_\mathrm{flow}$ of the outflow (see
Fig.~\ref{fig:global}). The possibility that the outflowing matter be
blocked by the infalling material from the RLOF stream has not been
considered, because the hotspot will be elongated as a consequence of
stellar rotation (the accreting star is spun up and no longer
synchronised, as observed by \citealt{1996MNRAS.283..613M}). It is
therefore very likely that there is only a small overlap between the
stream and the outflow.

\begin{figure}[t]
  \centering\includegraphics[width=0.48\textwidth]{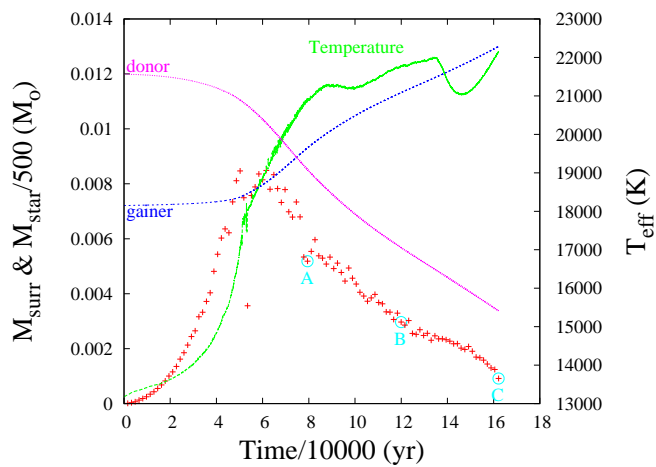}
  \caption{Red crosses: evolution of the mass surrounding the system
    (integrated up to the radius where the density of the outflow
    equals the ISM density; see text). The cyan circles (labelled with
    the model name) indicate the models selected for the simulations
    presented in this work. Long-dashed green line (right hand scale):
    evolution of the gainer-star surface temperature. Magenta dotted
    line and dashed blue line: evolution of the masses of the donor
    and gainer (in unit of $M_\odot/500$).}
  \label{fig:mass_surrounds}
\end{figure}

\begin{figure}[t]
  \centering\includegraphics[width=0.48\textwidth]{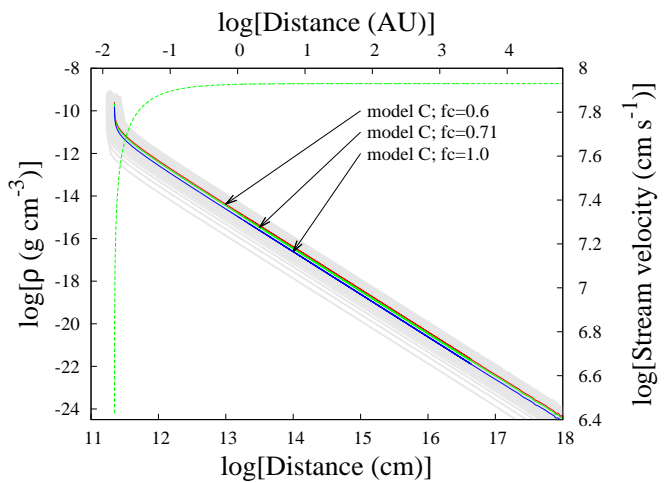}
  \caption{Density (left-hand scale, solid lines) and velocity
    (right-hand scale, dashed line) profiles of the outflowing
    material (similar -- but with different parameters -- to Fig.~9 of
    Paper~I). Blue line: spherically-symmetric and isotropic outflow
    corresponding to a covering factor of $f_\mathrm{c}=1$ (see text,
    Sect.~\ref{sec:rho}); green lines: ejection confined in a disc of
    opening angle $90^{\circ}$ ($f_\mathrm{c} = 0.71$); red line:
    ejection confined in a disc of opening angle $74^{\circ}$
    $(f_\mathrm{c}=0.6)$; The solid light-grey lines are the density
    profiles for $f_\mathrm{c} = 0.71$ at 100 different times
    uniformly distributed during the non-conservative phase.}
  \label{fig:density}
\end{figure}

By computing the density profile of the circumstellar material at
different stages during the evolution, we can estimate the total
amount of mass potentially detectable. Figure~\ref{fig:mass_surrounds}
shows the evolution of this quantity during the non-conservative
phase. We see that the total amount of detectable mass surrounding the
binary is maximal before the reversal of the mass ratio (mass ratio
$q=1$ at $t\sim 75\,000$~yr: the beginning of the Algol phase). The
gainer surface temperature mainly evolves during the rapid phase of
mass transfer and reaches a plateau when the mass ratio is reversed.

For our radiative-transfer simulations, we selected three
configurations (see Table~\ref{tab:char}): (i) right after the
mass-ratio reversal ($t = 80\,000$~yr; hereafter model A); (ii) at $t
= 120\,000$~yr (model B); (iii) at the end of the non-conservative
phase ($t = 162\,000$~yr; model C, our default configuration). In the
density profiles of models A and B, the densities in the ionised
region close to the system become so high that an increasing amount of
the starlight is transformed to recombination lines and continua that
are not explicitly calculated by \textsc{Cloudy}. Indeed, the model
atoms used by \textsc{Cloudy} account for a finite number of energy
levels and emission from recombinations to higher levels cannot be
included in the spectrum. As a result, a significant fraction of the
energy released in these recombinations and in the subsequent cascades
to the corresponding ground states are missing in the emergent
spectrum. In these conditions, it is difficult to predict observable
quantities, and we decided to use the density profile of the model C
for the models A and B while keeping the hotspot and stellar
parameters given by the evolutionary model.  Therefore, the
simulations corresponding to cases~A and B should be considered as
exploring the sensitivity of our results to the luminosity of the
hotspot, which is the property differing the most between cases~A, B
and C (Table~\ref{tab:char}). The sensitivity to the circumstellar
shell mass (which was the initial goal for considering cases~A, B and
C -- but which could not be fulfilled for the reason explained above)
may instead be probed by varying the dust-to-gas ratio $\Phi$. As
discussed in Sect.~\ref{sec:dust}, a simulation for case~C has been
run with 200 times more dust than standard (model labelled
'C\_hd\_0.71' in Table~\ref{tab:simu}). This case thus mimicks a
situation with a thicker dust shell.

Figure~\ref{fig:density} shows the density profile of the outflow for
different covering factors ($f_{\mathrm{c}}$) for model C. The profile
is cut at the distance $r_\mathrm{cut}$, where the density drops below
the ISM value. This occurs at $6.08 \times 10^{17}$~cm ($\approx
40\,600$~AU or 2.5$\times10^5~A_\mathrm{orb}$, where $A_\mathrm{orb}$
is the orbital separation) from the gainer for $f_\mathrm{c}=0.71$. We
do not take into account any bow-shock formation at this
location. Note that when the systemic mass loss stops, within 300
years the material disperses and the density profile drops everywhere
below the ISM density ($\approx$\,10$^{-24}$~g~cm$^{-3}$) because of
the very high velocity of the line-driven outflow
(Fig.~\ref{fig:density}).

\subsection{Spiral structure}
\label{sec:spiral}

Due to the orbital motion and the fact that the outflow comes from a
small area on the star, we must expect a spiral structure to form
close to the star. The width of the spiral arm increases as a function
of the distance from the star (Fig.~\ref{fig:global}) and eventually
the spiral arm will overlap itself so that the spiral structure fades
to give rise to a disc instead. Using an opening angle for the outflow
of 90$^{\circ}$ (standard $\theta_{\mathrm{flow}}$ configuration
corresponding to $f_{\mathrm{c}}=0.71$) and the velocity profile shown
in Fig.~\ref{fig:density}, we estimate the distance at which the
spiral fades to be $\approx3.6$~AU (22~$A_\mathrm{orb}$).  This
distance is reached by the outflow on the time scale of an orbital
period. At larger distances, our assumption of a smooth density
profile is thus sufficiently accurate. At smaller distances, we expect
deviations due to the spiral structure. In these regions, the density
will be the highest when the line of sight points towards the hotspot
and will decrease as the orbital motion moves the hotspot out of
sight. Ultimately, right before the hotspot points again toward the
observer, the density close to the star reaches a minimum as only the
material that has been launched orbital periods ago is seen far away
from the star. To mimic this effect and investigate any variability
due to the orbital motion, we perform some simulations with a reduced,
constant density profile close to the star (in the following, we call
these simulations `plateau' and label them with a `p' in the name).
Considering the orbital motion ($P_\mathrm{orb}=8\fd33$ at the end of
the non-conservative phase), we can estimate the density of the
outflow ahead of the hotspot. Given the travel time (one orbital
period) and the flow velocity, we can estimate the distance travelled
by the material, $r_\mathrm{pl}$, as well as the density at that
location, $\rho_\mathrm{pl}$. As a first approximation, we adopt a
density profile where the density is equal to $\rho_\mathrm{pl}$ up to
$r_\mathrm{pl}$ and then follow the standard (unmodified) density
profile. For the model with $f_\mathrm{c}=0.71$, we retrieve a radius
of $r_\mathrm{pl} = 6.05 \times 10^{13}$~cm ($\approx$\,4.05~AU or
25~$A_\mathrm{orb}$) and a corresponding density of
$\rho_\mathrm{pl}=9.44\times10^{-17}$~g\,cm$^{-3}$.

\subsection{Chemistry of the outflow: gas composition and
  dust formation}
\label{sec:dust}

One important point for the radiative-transfer analysis is the
composition of the gas, which we assume to be that of the donor as
listed in Table~\ref{tab:composition}. The main signature is the
depletion of carbon and enhancement of nitrogen during the
non-conservative phase (since the mass transfer reveals the inner
layers of the giant star).

Although grains can be useful tracers of the cool outermost regions of
the outflow, it is not clear whether or not they can form. Indeed, the
escaping material cools as it leaves the gainer's
vicinity. Eventually, the temperature drops below the condensation
temperature, allowing for the formation of grains. However, UV and
X-ray photons from the hotspot and star (e.\,g.,
\citealt{2008A&A...477L..37I}) may prevent grain
formation. Surprisingly, grains are found in the surroundings of
Wolf-Rayet stars \citep{2008A&A...486..971B}, which are also emitters
of hard radiation.

As there are no observational constraints on the chemical composition
and optical properties of (hypothetical) dust grains around Algol
systems, we use the standard ISM model implemented in
\textsc{Cloudy}. However, we exclude carbonaceous grains, because the
evolutionary stages of both stars imply C/O~$<1$ (see
Table~\ref{tab:composition}) so that we can assume that most carbon
atoms in the outflow are locked up in CO molecules. For the slope of
the size distribution of the remaining silicate grains we use the
expression suggested by \citet{1977ApJ...217..425M},
\begin{equation}
  \label{eq:grsize}
  \frac{\mathrm{d}n_{\mathrm{gr}}}{\mathrm{d}a} \propto a^{-3.5},
\end{equation}
where $n_{\mathrm{gr}}$ is the number density of grains and $a$ the
grain size. The size distribution is discretised in ten bins, which is
the maximum number of size bins allowed by Cloudy, ranging from 0.005
to 0.25~{\textmu}m, with a ratio
$a_{\mathrm{max}}/a_{\mathrm{min}}\approx 1.48$ in each bin, where
$a_{\mathrm{max}}$ and $a_{\mathrm{min}}$ are the upper and lower size
limit of the bins, respectively. Within each bin, the optical
properties of the dust grains are averaged over the respective size
interval. For our simulations, the chosen number of bins and their
width and location have very little impact on the results. Varying the
number and size of the bins only results in a deviation of $\la2$\% in
the flux ratios and magnitudes. \cite{2008A&A...486..971B} evaluate
the impact of the lower cut-off on the grain size (0.005~{\textmu}m in
our case) and find that only the flux blueward of 8.3 {\textmu}m is
altered, which has no influence on WISE bands 3 and 4 ($>10$
{\textmu}m) that will be used in Sect.~\ref{sec:wise}. Similarly, we
ran simulations with different grain sizes and found that the region
of the spectra below $\sim$6~{\textmu}m does not depend on grain size
so that 2MASS fluxes ($<5$~{\textmu}m) are not affected.

The optical properties of the grains are calculated by use of Mie
theory as described by \citet{2004MNRAS.350.1330V}. As the gainer star
is of spectral type B, its radiation will probably interfere with the
formation of dust grains, and we cannot expect the dust-to-gas ratio
($\Phi$) to be as high as in the ISM. Instead, we use the results of
\citet{2008A&A...486..971B}, who find dust in the surroundings of WN
stars. From their best-fit models we derive values of $\Phi$ up to
$2\times10^{-5}$, which we adopt in this work for our standard dusty
configuration (models labelled `ld' for low-dust). To investigate the
impact of the dust-to-gas ratio, we also computed a model with $\Phi =
4\times10^{-3}$ (model labelled `hd' for high-dust).

We do not explicitly calculate the destruction of dust grains caused
by the high temperatures close to the star. Instead, we use the
\textsc{Cloudy} code to derive a temperature profile for a pure-gas
model (models labelled `df' for dust-free) and then define a dust-free
region around the star where the temperature exceeds the dust
sublimation temperature ($\approx$\,1400~K for silicate dust, see
\citealt{1991eua..coll..341M, 1993ApJ...402..441L}). The circumstellar
material will be dust-free up to a transition radius $r_{\mathrm{tr}}$
(radius at which the temperature goes below the condensation
temperature).

\subsection{Stellar parameters}
\label{sec:setup_cloudy}

\begin{table}
  \caption{Parameters for the three models; Model A: mass ratio
    reversal; Model B: intermediate case; Model C: end of
    non-conservative mass transfer (for this case, three hotspot
    temperatures are investigated).}
  \label{tab:char}
  \centering
  \begin{tabular}{@{}l@{ }c@{ }c@{ }c@{}}
    \hline
    \hline\\
    \textbf{Model A} & Gainer & Donor & Hotspot \\ \\
    \hline\\
    Radius (R$_{\odot}$)   & 3.87     & 5.73    & 0.14 (0.19)\\
    Mass (M$_{\odot}$)       & 4.65     & 4.26    & ---        \\
    Luminosity (L$_{\odot}$) & 2\,496   & 54      & 7         \\
    Temperature (K)        & 20\,840  & 6\,555  & 35\,000 \\ 
    $P_{\mathrm{orb}}$ && 2.3 d&\\
    $A_{\mathrm{orb}}$ && 15.5 R$_{\odot}$&\\ \\
    \hline
    \hline\\
    \textbf{Model B} & Gainer & Donor & Hotspot \\ \\
    \hline\\
    Radius (R$_{\odot}$)   & 3.13     & 6.19    & 0.24 (0.26) \\
    Mass (M$_{\odot}$)       & 5.66     & 2.83    & ---     \\
    Luminosity (L$_{\odot}$) & 1\,924   & 38      & 20      \\
    Temperature (K)        & 21\,780  & 5\,784  & 35\,000  \\ 
    $P_{\mathrm{orb}}$  &&3.3 d&\\
    $A_{\mathrm{orb}}$  &&19.0 R$_{\odot}$&\\ \\
    \hline
    \hline\\
    \textbf{Model C} & Gainer & Donor & Hotspot \\ \\
    \hline\\
    Radius (R$_{\odot}$)   & 3.22     & 9.55    & 0.87 (0.63) (C\_xx\_0.71\_cold)\\
                           &  &  & 0.64 (0.63)  (C\_xx\_0.71\_inter)\\ 
                           &  &  & 0.49 (0.63) (C\_xx\_0.71\_hot)\\\\
    Mass (M$_{\odot}$)       & 6.50     & 1.68    & ---        \\
    Luminosity (L$_{\odot}$) & 2\,195   & 63      & 137        \\
    Temperature (K)        & 22\,150  & 5\,297  & 30\,000 (C\_xx\_0.71\_cold)\\
       &  &  & 35\,000 (C\_xx\_0.71\_inter)\\ 
       &  &  & 40\,000 (C\_xx\_0.71\_hot)\\
    $P_{\mathrm{orb}}$  &&8.3 d&\\
    $A_{\mathrm{orb}}$  &&34.9 R$_{\odot}$&\\ \\ \hline
  \end{tabular}
  \tablefoot{The different temperatures of the hotspot correspond to
    different simulations labelled (cold, inter, hot). `xx' stands for
    `ld' or `hd' and designates different dust-to-gas ratios (see
    text). $P_{\mathrm{orb}}$ and $A_{\mathrm{orb}}$ are the orbital
    period and separation, respectively. Refer to Table~\ref{tab:simu}
    for further explanations. The hotspot radius is calculated using
    the black body relation (Sect.~\ref{sec:thotspot}) and the
    estimate given by Eq.~\ref{eq:radius_hot} is shown for comparison
    in parenthesis. The density profile of model C is used for all
    three models (see Sect.~\ref{sec:rho}).}
\end{table}

\subsubsection{Luminosities}
\label{sec:luminosity}

The system is composed of the donor star (a late G-type red giant),
the gainer star (a B-type dwarf), the hotspot and the material
escaping from the hotspot (Fig.~\ref{fig:global}). The properties of
the system (masses, radii, stellar and hotspot luminosities, stellar
temperatures; \citealt{2013A&A...557A..40D}) are computed with the
\textsc{Binstar} code and given in Table~\ref{tab:char} for the three
selected models.

We can estimate the size of the hotspot ($r_{\mathrm{hs}}$) as being
identical to the size of the impacting stream ($r_{\mathrm{stream}}$),
which can in turn be identified with the size of the stream at the
Lagrangian point $\mathcal{L}_{1}$. According to
\cite{1985ibs..book.....P}, it can be approximated by
\begin{equation}
  \label{eq:radius_hot}
  r_{\mathrm{hs}} \sim r_{\mathrm{stream}} = \frac{1}{2}
  \frac{c_{\mathrm{s}}}{\Omega_{\mathrm{orb}}},
\end{equation}
where $c_{\mathrm{s}}$ is the sound speed of the material leaving the
donor star at $\mathcal{L}_{1}$ and $\Omega_{\mathrm{orb}}$ is the
orbital angular velocity.

The effective luminosity felt by the material escaping the system
depends on its location along the spiral. In the cone with opening
$\theta_{\rm flow}$ above the hotspot (Fig.~\ref{fig:global}), the
escaping material is irradiated by both the gainer and the
hotspot\footnote{We assume that the donor star does not contribute to
  the irradiation of the outflowing material, because its luminosity
  is much lower than that of the gainer (see
  Table~\ref{tab:char}). This, however, may not remain true in Algols
  with a fainter gainer of spectral type A.}.  But outside this cone,
the hotspot luminosity does not contribute any more. We account for
this effect by using an effective hotspot luminosity, corresponding to
the actual value reduced by a geometrical factor $\gamma$.  This
geometrical factor is obtained (Appendix~\ref{sec:flux}) by assuming
that the hotspot luminosity is diluted over a strip-shaped surface
encircling the star (blue band in Fig.~\ref{fig:geometry}). The
effective flux from that surface reads
\begin{equation}
    F_\mathrm{eff}=F_\mathrm{g}+\gamma F_\mathrm{hs},
\end{equation}
where $F_{\mathrm{g}}=L_*/\mathcal{A}_*$ and
$F_{\mathrm{hs}}=L_\mathrm{hs}/\mathcal{A}_\mathrm{hs}$ are the
surface fluxes of the gainer star and the hotspot, respectively (see
Appendix~\ref{sec:flux} for details on the calculation of the factor
$\gamma$). With the quantities given in Table~\ref{tab:char}, we find
$\gamma\sim0.05$ for model C.

The use of a time-independent configuration, with an effective flux
that constantly irradiates all the surrounding material, is justified
by the fact that the cooling time-scale (as computed by
\textsc{Cloudy}; see Fig.~\ref{fig:temp}) becomes longer than the
orbital period at distances $r\approx 3.8\times 10^{12}$~cm (0.25~au)
for model C.  Due to the high outflow velocity (up to
900~km\,s$^{-1}$; Fig.~\ref{fig:density}), the time needed for the
flow to reach that distance is short
($\approx13\fd{93}=1.67P_\mathrm{orb}$ at the end of the
non-conservative phase). Therefore, the outflow properties (like
temperature, emission spectrum, ...) will not change too much over one
orbital period, and several phases of irradiation (as the material
passes in front of the hotspot) can be averaged into a constant
irradiation.

The spectrum of the gainer star is obtained by interpolating in the
`TLUSTY' grid of \citet{2007ApJS..169...83L} to match the effective
temperature given in the first column of Table~\ref{tab:char}. The
donor-star spectrum is taken as a black-body spectrum (see upper panel
of Fig.~\ref{fig:SED}). For all the components constituting the total
incident spectrum (donor, gainer, hotspot), we use the metallicities
listed in Table~\ref{tab:composition}. The donor has by definition the
same abundances as the material ejected and, for the selected models,
the gainer star is entirely covered by the donor's material.

\subsubsection{Distance}
\label{sec:distance}

\begin{figure}[t]
  \centering
  \includegraphics[width=0.49\textwidth]{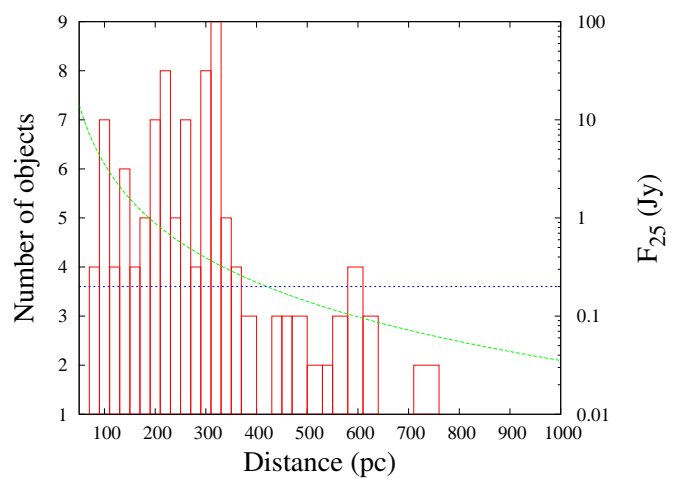}
  \caption{Distance distribution for observed Algols. The sample of
    Algols is from \cite{2004A&A...417..263B}. Parallaxes are from the
    Hipparcos catalogue \citep{1997ESASP1200.....P}. The dashed green
    curve represents the flux at 25~{\textmu}m (right-hand scale) as a
    function of distance for the dusty simulation
    C\_ld\_0.71\_inter. The flux is integrated over the whole
    nebula. The dotted blue line is the IRAS 5$\sigma$ limit at
    25~{\textmu}m. The WISE 5$\sigma$ limit for band $W4$ is
    2600~{\textmu}Jy and is always below the dashed green curve.}
  \label{fig:distance}
\end{figure}

To relate absolute fluxes (from the models) with predicted apparent
magnitudes for e.g., the infrared excesses, the distance to the system
needs to be fixed.  We choose a typical distance from the distribution
of measured parallaxes for Algols.  In Fig.~\ref{fig:distance}, we
present the distance distribution for 152 objects [join entries in the
  \cite{2004A&A...417..263B} and Hipparcos \citep{1997ESASP1200.....P}
  catalogues], which correspond to about 10\% of all Algols in the
\cite{2004A&A...417..263B} catalogue. To evaluate which among these
systems might be detectable by IRAS, Fig.~\ref{fig:distance} also
presents the 25~{\textmu}m flux of the dusty simulation
C\_ld\_0.71\_inter, integrated over the whole nebula, as a function of
distance (dashed green line). Only the infrared sources with a flux
larger than the IRAS 5$\sigma$ detection limit at 25~{\textmu}m (blue
horizontal dotted line) can be detected, so that a system with the
same IR emission as our model C\_ld\_0.71\_inter must be within 400~pc
of the Sun for its IR excess to be detected, and this includes most of
the Algols with a known parallax. Based on this result, a typical
distance of 300~pc is adopted to convert model fluxes into apparent
fluxes and magnitudes, as will be listed in Sects.~\ref{sec:results}
and \ref{sec:wise}. To give an idea, $\beta$~Lyr\ae{} is located at a
distance of 294~pc while \object{W Ser} is estimated to lie at
1\,428~pc \citep[Hipparcos catalogue,][]{2007A&A...474..653V}. Of
course, the apparent fluxes of the model SEDs can be easily rescaled
for comparison with a system at a different distance.

\subsubsection{The temperature of the hotspot}
\label{sec:thotspot}

To assess the dependence of our results on the SED of the hotspot, we
use three different hotspot temperatures $T_\mathrm{hs}$ in model C
with $f_\mathrm{c}=0.71$ (see Tables~\ref{tab:char} and
\ref{tab:simu}).

In the first model (C\_xx\_0.71\_cold), we assume that the hotspot
emits with a B-star spectrum at a temperature of 30\,000~K\footnote{In
  our models, the hotspot is a region of the star's atmosphere with a
  higher temperature and we therefore use a synthetic star spectrum
  instead of a black-body spectrum.}. In the second model
(C\_xx\_0.71\_inter), we assume a temperature of 35\,000~K and use an
O-star spectrum (also interpolated from the TLUSTY grid). In the third
model (C\_xx\_0.71\_hot), we use an O-star spectrum with a temperature
of 40\,000~K. These three temperatures are based on observed hotspot
values in Algols with similar properties (stellar masses and orbital
periods; \citealt{2011A&A...528A..16V}), and the corresponding results
are presented in Sect.~\ref{sec:hotspot_configuration}. These three
models allow to clearly disentangle the effect of the hotspot
properties on the emergent SED.

In order to satisfy the Stefan-Boltzmann law $L_{\mathrm{hs}}=\sigma
T_{\mathrm{hs}}^4 \mathcal{A}_{\mathrm{hs}}$ (cf.\ Eq.~\ref{ap:aire}),
the radius of the hotspot is treated as a free parameter
($L_{\mathrm{hs}}$ and $T_{\mathrm{hs}}$ are fixed), and compared to
the estimate given by Eq.~\ref{eq:radius_hot} (see
Table~\ref{tab:char}). Although Eq.~\ref{eq:radius_hot} can only give
a rough approximation of the radius of the hotspot, we see a good
match between this later value and the one imposed by the
Stefan-Boltzmann law.

\subsection{Setup of \textsc{Cloudy} and \textsc{Skirt}}
\label{sec:codes}

We use version \oldstylenums{13.02} of \textsc{Cloudy}, last described
by \citet{2013RMxAA..49..137F}, to calculate the ionisation of the gas
close to the gainer star and the emission from the dust grains in the
outflow at larger distances. \textsc{Cloudy} calculates the thermal
and ionisation balance with a simultaneous treatment of radiative
transfer. For the latter, an escape probability formalism is used. In
our simulations, the photon energy scale is discretised with a
resolution of $\lambda/\Delta\lambda=2\,000$ down to $\lambda =
0.152$~nm\footnote{$\lambda/\Delta\lambda=333$ between
  \textsc{Cloudy}'s high-energy limit at 12.4~fm and
  0.152~nm.}. \textsc{Cloudy} calculates the strengths of $\ga10^5$
spectral lines. We found that the number of energy levels that are
used for the \ion{Fe}{ii} ion can have a strong impact on the
resulting temperature profile and thus on the transition radius
$r_\mathrm{tr}$, and also on the emitted spectrum. Therefore, we used
the detailed \ion{Fe}{ii} model that was developed by
\citet{1999ApJS..120..101V} and implemented in \textsc{Cloudy}. To
define the spatial step size, \textsc{Cloudy} uses adaptive
logic. However, for the models C\_df\_0.71\_hot, C\_ld\_0.71\_hot,
C\_df\_1.00\_inter, and C\_ld\_1.00\_inter, we had to define a lower
limit for the step size, of the order of $10^8$~cm, to reach
convergence.

The ejection of matter triggered by the hotspot is assumed to occur
preferentially in the orbital plane and is assumed to be seen
edge-on. In our spherical 1D simulations with \textsc{Cloudy}, we take
that into account by adjusting the covering factor (equivalent to the
factor $f_\mathrm{c}$ introduced in Sect.~\ref{sec:rho},
Eq.~\ref{eq:rho}). In Appendix~\ref{ap:opening_angle} we estimate the
covering factor (or, equivalently, the opening angle) based on the
assumption of isotropic radiation from the hotspot. To study the
effect of the choice of the opening angle on the results, we performed
simulations with different covering factors (the three digits in the
label of each simulation: `x.xx').

To assess the systematic uncertainties due to the 1D geometry and the
simplified approach used by \textsc{Cloudy} for the radiative
transfer, we use the \textsc{Skirt} code, which simulates the general
3D continuum transfer for dust, including absorption, multiple
anisotropic scattering, and thermal emission
\citep{2003MNRAS.343.1081B, 2011ApJS..196...22B,
  2015A&C.....9...20C}. The 3D dust-distribution is represented by a
disc that is centred on the gainer star and truncated at its inner
edge at the dust condensation radius. The thickness of the disc
increases linearly with distance as defined by the opening angle
$\theta_\mathrm{flow}$ (see Appendix~\ref{ap:opening_angle}), and the
density inside the disc is assumed to be a function of distance only
(see Fig.~\ref{fig:density}). The spatial grid consists of 200
concentric shells. The density is set to zero in the directions where
the polar angle is outside the range
$\pi/2-\theta_\mathrm{flow}/2\leq\theta\leq\pi/2+\theta_\mathrm{flow}/2$,
and the extent of the grid cells is adapted accordingly. The positions
$r_i$ of the grid cells along the radial coordinate follow a power-law
distribution of the form
\begin{equation}
 \label{eq:grid}
 r_{i+1}=r_\mathrm{tr}+\frac{1-R^{i/(N-1)}}{1-R^{N/(N-1)}}
 (r_\mathrm{max}-r_\mathrm{tr}),
\end{equation}
where $N>i\geq0$ and $r_0=0$. $r_\mathrm{tr}$ is the dust formation
radius (212~AU for the model C\_ld\_0.71\_inter;
Sect.~\ref{sec:temperature_profile}), $N=200$ the number of grid
points, and $r_\mathrm{max}$ is chosen in such a way that the density
drops below the ISM density at the outer edge of the grid. $R=800$ is
approximately the ratio of the sizes of the outermost and innermost
grid cells, $(r_N-r_{N-1})/(r_2-r_1)$. The functional form of
Eq.~\ref{eq:grid} is motivated by the steeply decreasing density
profile. The ratio of the widths of two adjacent grid cells is held
constant. This implies that the width of the grid cells increases from
the centre towards the outer edge. Therefore, the high-density regions
close to the centre are covered by a larger number of cells than the
outer low-density regions. This allows us to save computing time.  The
first term on the right-hand side of Eq.~\ref{eq:grid} is introduced
because there is a dust-free region around the centre, which does not
have to be resolved in the \textsc{Skirt} simulations. The composition
and size distribution of the dust are the same as in the
\textsc{Cloudy} models. In the radiative-transfer Monte-Carlo method
used in \textsc{Skirt}, the radiation field is represented by discrete
photon packages. We used $10^{6}$ photon packages per wavelength in
our simulations. Our wavelength grid consists of 1\,000 wavelength
points with logarithmic spacing between 1~nm and 500~{\textmu}m, which
allows convergence within a reasonable computation time.

The different parameters used in the simulations performed with
\textsc{Cloudy} and \textsc{Skirt} are summarised in
Table~\ref{tab:simu}.
\begin{table*}
  \centering
  \caption{Parameters of the simulations performed with the \textsc{Cloudy}
    and \textsc{Skirt} codes.}
  \begin{tabular}{|@{\,}c@{\,}||@{\,}c@{\,}|@{\,}c@{\,}|@{\,}c@{\,}|@{\,}c@{\,}|@{\,}c@{\,}|@{\,}c@{\,}|@{\,}c@{\,}|}
    \hline
    & \multicolumn{5}{c|}{} & \multicolumn{2}{|c|}{} \\
    \textbf{code} & \multicolumn{5}{c|}{\textbf{\textsc{Cloudy}}} &
    \multicolumn{2}{c|}{\textbf{\textsc{Skirt}}} \\
    & \multicolumn{5}{c|}{} & \multicolumn{2}{|c|}{} \\
    \hline
    \hline
    &\multicolumn{5}{c|}{} & \multicolumn{2}{|c|}{} \\
    {\boldmath{$f_\mathrm{c}$ or $\theta_{\mathrm{flow}}$}} &
    \multicolumn{5}{c|}{0.60 / 0.71 / 1.00}
    & \multicolumn{2}{c|}{$90^\circ$} \\
    & \multicolumn{5}{c|}{} & \multicolumn{2}{|c|}{} \\
    \hline
    &\multicolumn{2}{c|}{} & \multicolumn{3}{|c|}{} &
    \multicolumn{2}{c|}{} \\
    \textbf{density} & \multicolumn{2}{c|}{plateau} &
    \multicolumn{3}{c|}{normal} & 
    \multicolumn{2}{c|}{normal} \\
    &\multicolumn{2}{c|}{} & \multicolumn{3}{c|}{} &
    \multicolumn{2}{c|}{} \\
    \hline
    &  &  &
    & \multicolumn{2}{c|}{} & \multicolumn{2}{c|}{} \\
    \textbf{chemistry} & gas & gas + dust &
     gas & \multicolumn{2}{c|}{gas + dust} & \multicolumn{2}{c|}{dust} \\
    &  & & & \multicolumn{2}{c|}{} &
    \multicolumn{2}{c|}{} \\ 
    \hline
    &&&&&&&\\
    {\boldmath{$\Phi$}} & --- & $2\times10^{-5}$ &
    --- & $2\times10^{-5}$ & $4\times10^{-3}$ & $2\times10^{-5}$  &
    $4\times10^{-3}$ \\
    &&&&&&&\\
    \hline
    &&&&&&&\\
    \textbf{label} & C\_df\_p\_0.71 &  &
    \parbox{2.5cm}{\centering A\_df\_0.71
      \\ B\_df\_0.71 \\ C\_df\_0.71\_cold \\ C\_df\_x.xx\_inter \\ C\_df\_0.71\_hot}
    &\parbox{2.5cm}{\centering A\_ ld\_0.71 \\ B\_ld\_0.71
      \\ C\_ld\_0.71\_cold \\ C\_ld\_x.xx\_inter \\ C\_ld\_0.71\_hot} & 
    C\_hd\_0.71 & C\_SK\_ld  & C\_SK\_hd \\
    &&&&&&&\\
    \hline
  \end{tabular}
  \tablefoot{`plateau' (models labeled `p') indicates that the
    simulation has a constant density in the inner part (see text,
    Sect.~\ref{sec:spiral}). $\Phi$ is the dust-to-gas ratio,
    $f_\mathrm{c}$ the covering factor (used for the \textsc{Cloudy}
    simulations) and $\theta_{\mathrm{flow}}$ the opening angle (used
    in the \textsc{Skirt} simulations to define a 3D dust
    distribution). In all models, the hotspot temperature is set to
    $T_\mathrm{hs}=35\,000$~K (i.\,e. `inter'), except for the models
    with $f_\mathrm{c}=0.71$, for which we applied three different
    hotspot temperatures (`cold': $T_\mathrm{hs} = 30\,000$~K;
    `inter': $T_\mathrm{hs} = 35\,000$~K; `hot': $T_\mathrm{hs} =
    40\,000$~K) with different spectra (see
    Sect.~\ref{sec:thotspot}). `A', `B' and `C' refer to the
    evolutionary phase (see Fig.~\ref{fig:mass_surrounds} and
    Table~\ref{tab:char}), `x.xx' is a three-digit number for the
    covering factor, `df' stands for dust-free, `ld' for low dust
    (i.\,e., $\Phi = 2 \times 10^{-5}$), `hd' for high dust (i.\,e.,
    $\Phi = 4 \times 10^{-3}$).}
  \label{tab:simu}
\end{table*}

\section{Results}
\label{sec:results}

\subsection{Temperature and emissivity profiles}
\label{sec:temperature_profile}

\begin{figure}[t]
  \centering \includegraphics[width=0.48\textwidth]{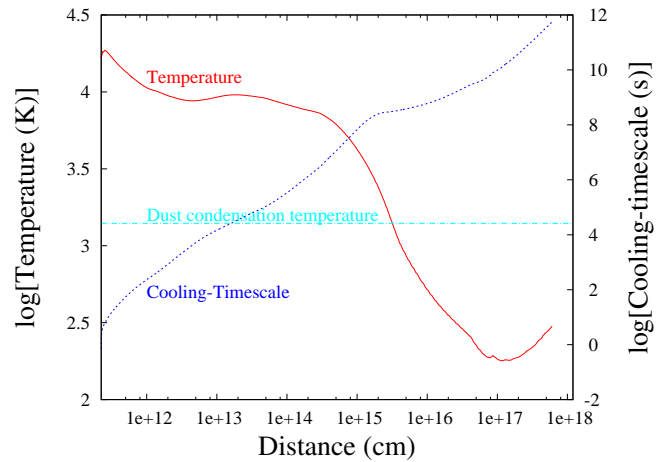}
  \caption{Temperature profile (solid red line) and cooling time-scale
    (dotted blue line; see text) of the material surrounding the star
    as a function of distance to the star (model
    C\_df\_0.71\_inter). The dot-dashed cyan line represents the dust
    condensation temperature (1\,400~K).}
  \label{fig:temp}
\end{figure}
The first quantity of interest is the outflow-temperature profile,
because dust can only form if it falls below the condensation
temperature.  Figure~\ref{fig:temp} shows the temperature profile for
the model C\_df\_0.71\_inter (see Table~\ref{tab:simu}). The
temperature drops below the dust condensation temperature at
$r_\mathrm{tr}=3.2\times10^{15}$~cm ($\approx 212$~AU), which is small
compared to the extent of the outflow $r_\mathrm{cut}\approx 6.1
\times10^{17}$~cm (see Sect.~\ref{sec:rho}). Of course, this value
depends on the covering factor and incident spectrum.

The temperature profile shown in Fig.~\ref{fig:temp} is a decreasing
function of distance in the inner regions, with temperatures dropping
from $\approx$\,17\,000~K to $\approx$\,180~K in the first $ \approx 1
\times 10^{17}$~cm ($\approx$\,6\,700~AU) from the star's
centre. Then, due to the steeply decreasing density, the temperature
increases again up to about 300~K.

Two other interesting quantities to follow are the profiles of the
volume emissivity at some specific wavelengths (especially for the
investigation of dust), like the central wavelengths for some
widely-used IR-photometry systems (2MASS, WISE;
\citealt{2003yCat.2246....0C, 2012yCat.2311....0C}), and the
ionisation fraction profiles of some ions that are observed in the
emission-line spectra of many Algol-related systems (\ion{Fe}{ii},
\ion{Si}{iv}, etc...). The ionisation fraction for an ion \ion{X}{z}
is defined as $w_{\ion{X}{z}}=n_{\ion{X}{z}}/n_{\mathrm{X,total}}$,
where $n_{\mathrm{X,total}}$ is the total number density of
element \ion{X}{}. These profiles tell us where most of the flux comes
from.

\begin{figure}
  \centering \includegraphics[width=0.49\textwidth]{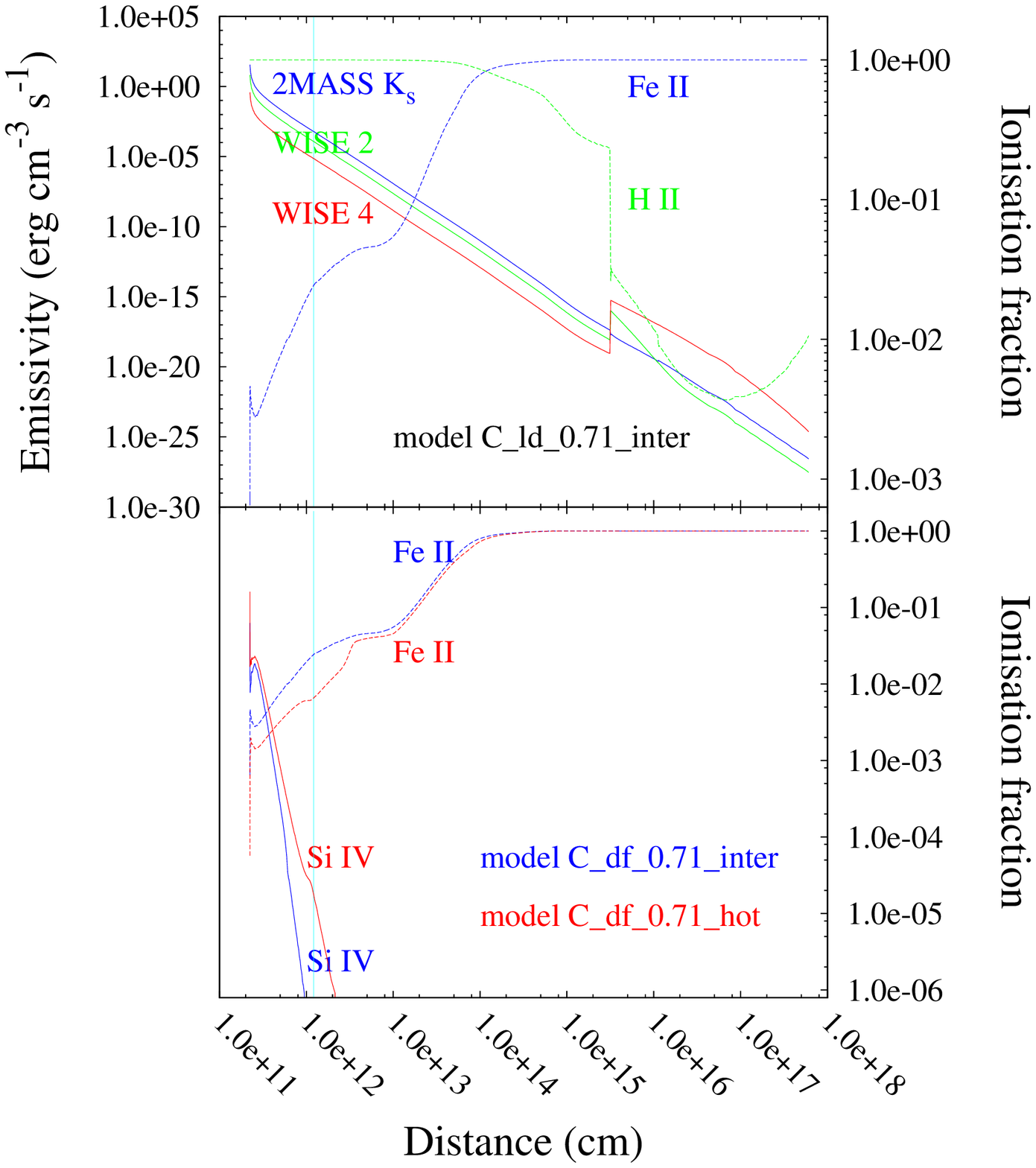}
  \caption{Top panel: Emissivity profiles (solid lines) and relative
    ionisation (dashed lines) for the simulation C\_ld\_0.71\_inter
    (blue solid line: 2MASS $K_{\mathrm{s}}$; green and red solid:
    WISE $W2$ and $W4$, respectively; green dashed: \ion{H}{ii}
    fraction; blue dashed: \ion{Fe}{ii} fraction). The vertical cyan
    line represents the Roche lobe radius for the gainer star. Bottom
    panel: \ion{Si}{iv} (solid lines) and \ion{Fe}{ii} fraction
    (dashed lines) for the high-temperature hotspot simulations (blue:
    model C\_df\_0.71\_inter; red: C\_df\_0.71\_hot).}
  \label{fig:emissivity}
\end{figure}

The top panel of Fig.~\ref{fig:emissivity} shows the emissivity
profiles of the 2MASS $K_{\mathrm{s}}$, WISE $W2$ and $W4$ bands for
the dusty simulation with $f_\mathrm{c}=0.71$ and $\Phi = 2\times
10^{-5}$ (model C\_ld\_0.71\_inter). Most of the flux comes from the
inner part of the system, but the infrared excess in these specific
bands comes from the outer dusty regions at
$r\ga2\times10^{15}$~cm. We indeed clearly see that $W4$ dominates
$W2$ and 2MASS $K_{\mathrm{s}}$ in this outer region, which is not the
case at smaller radii. In the outer region, the emissivity is low (of
the order of $10^{-15}$~erg\,cm$^{-3}$\,s$^{-1}$) but the total
emission, which is the integral over the volume, remains large. The
infrared excess in the $W4$ band is therefore visible as a `bump' in
the infrared part of the SED from $\sim$10~{\textmu}m towards longer
wavelengths (Fig.~\ref{fig:SED_IR}). Due to the rapidly decreasing
density profile, only the dust the closest to the star contributes to
the infrared excess, especially at the lowest wavelengths. For this
simulation, as well as for the dust-free one, the inner part is
totally ionised (\ion{H}{ii}, \ion{Fe}{iii} and higher). Although
\ion{Fe}{ii} is mainly present in the outer part of the system, the
formation of \ion{Fe}{ii} lines may partially happen close to the star
and present some line variability there (see
Sect.~\ref{sec:hotspot_configuration} and bottom panel of
Fig.~\ref{fig:emissivity}). In contrast, the contribution from
\ion{Fe}{ii} in the outer parts comes from a smooth, roughly
time-steady density distribution and is probably not variable. The
resulting observed line will be the sum of the two contributions
(variable inner part, steady outer part).

\subsection{Synthetic SEDs}
\label{sec:output_spectra}

In this section, we inspect the output spectra in the
different regimes, from ultraviolet to infrared, and the impact of the
different parameters.

\subsubsection{Optical-UV}
\label{sec:spectra_UV}

\begin{figure*}[t]
  \centering \includegraphics[width=1.\textwidth]{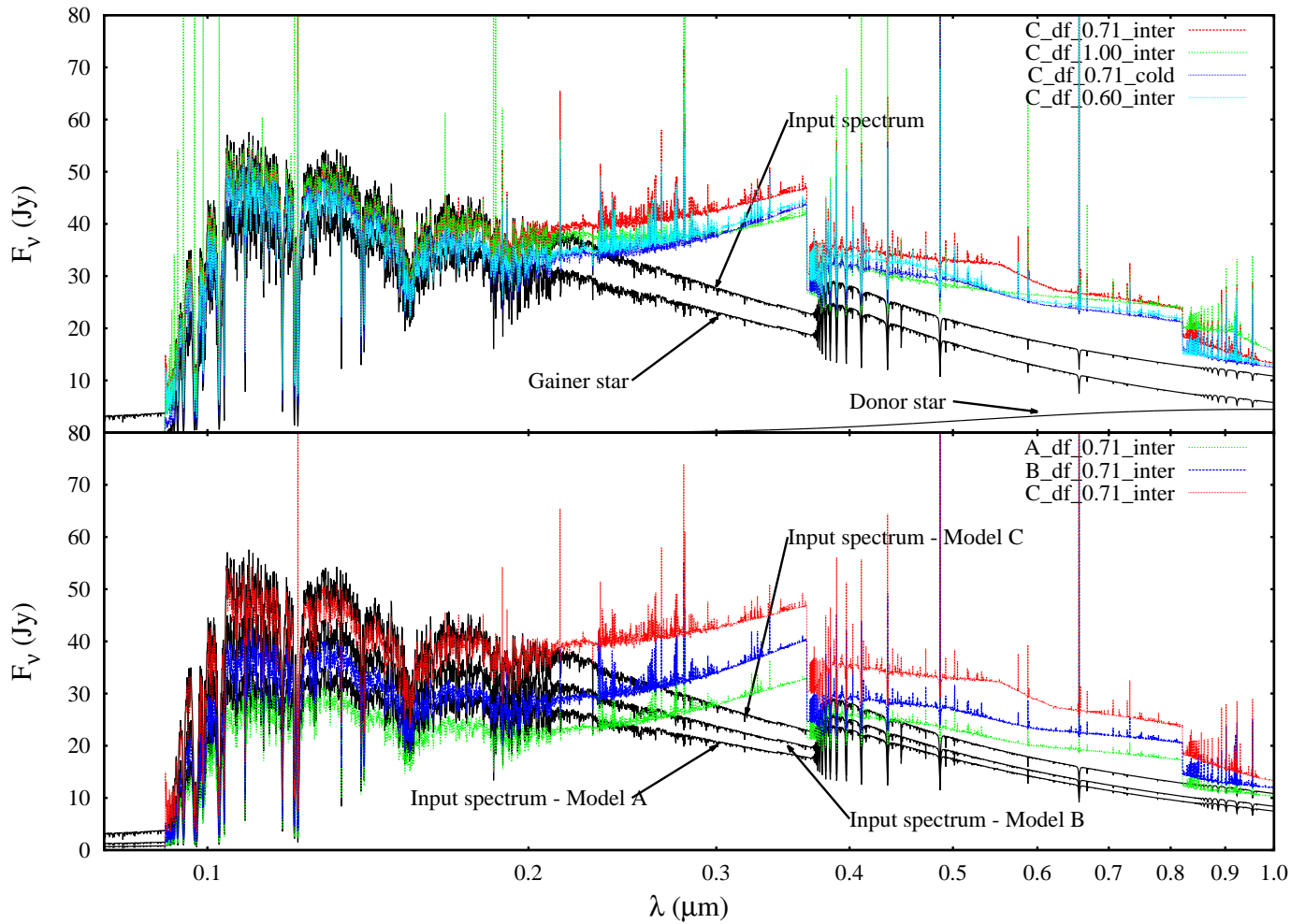}
  \caption{\textbf{Top:} Synthetic SEDs of the dust-free \textsc{Cloudy}
    models in the UV-optical regime. The input spectrum is the sum of
    the spectra of the gainer star, donor star, and hot spot with a
    temperature of 35\,000~K (the latter diluted according to the
    prescriptions of Appendix~B).  The fluxes are computed for a
    system distance of 300~pc.  \textbf{Bottom:} Same as top panel but
    for the different models A, B and C (at different evolutionary
    times). The input spectra are in black from bottom to top (A, B,
    C). }
  \label{fig:SED}
\end{figure*}

The top panel of Fig.~\ref{fig:SED} shows the emergent spectral energy
distributions (SEDs) for the dust-free models C\_df\_0.60\_inter,
C\_df\_0.71\_inter, C\_df\_0.71\_cold and C\_df\_1.00\_inter (the
corresponding SEDs for models that contain dust are nearly identical
in this spectral region and are therefore not shown). They all show
pronounced Balmer and Paschen recombination continua. The Balmer jump
in emission leads to observable diagnostics, as further discussed in
Sect.~\ref{sec:obs:stromgren}.

In the SEDs of the models C\_df\_0.71, we notice a slight absorption
in the UV up to a wavelength of $\sim0.2$~{\textmu}m (also seen in
Fig.~\ref{fig:SED_plateau} for the same model), which is mainly
attributable to photoionisation close to the star, where the density
is high. This has two consequences: i)~this absorption will quickly
disappear when the non-conservative phase will stop, because the
outflowing material will rapidly dissipate (see Sect.~\ref{sec:rho});
ii) the UV absorption should show a phase dependence owing to the
spiral structure of the outflow (this prediction is compared to
observations of W~Ser in Appendix~\ref{ap:WSer_UV}).

The bottom panel of Fig.~\ref{fig:SED} presents the SED for the
different models (A, B, C) with a covering factor $f_\mathrm{c}=0.71$
(corresponding to $\theta_{\mathrm{flow}} = 90^{\circ}$) and
$T_{\mathrm{hs}} = 35\,000$~K. Model C exhibits the strongest Balmer
and Paschen recombination continua and model A the most prominent UV
absorption. As discussed in Sect.~\ref{sec:obs:stromgren}, The
presence of the recombination continuum and emission lines, and the
absence of stellar absorption lines are not typical of Algols, but are
a common feature of symbiotic systems or planetary nebul\ae{}
\citep{1986syst.book.....K}.

\subsubsection{Far UV-continuum orbital modulation} 

\begin{figure}
  \centering
  \includegraphics[width=0.5\textwidth]{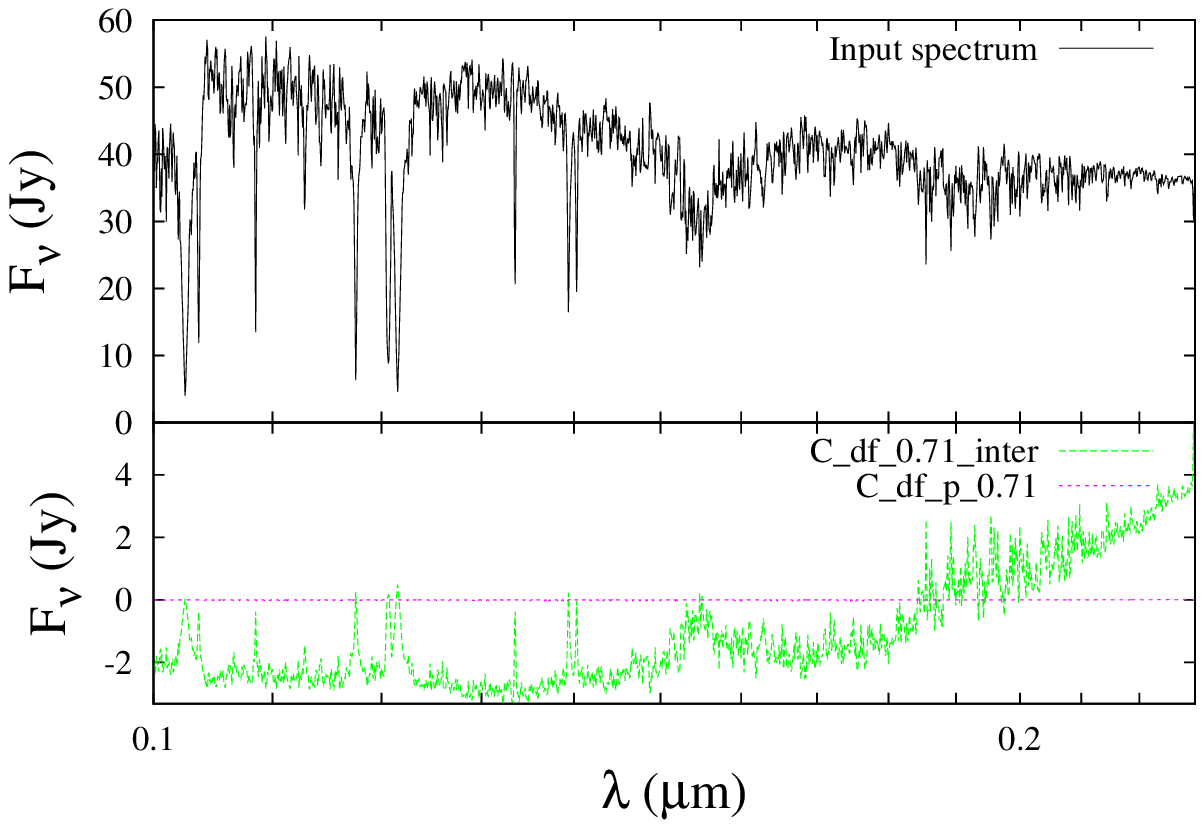}
  \caption{Synthetic SEDs of dust-free models in the UV regime. The
    solid black line in the upper panel represents the input stellar
    spectrum. The lower panel shows the difference between the
      output and input spectra for the two models identified by the
      line labels.}
  \label{fig:SED_plateau}
\end{figure}

The comparison of the results from the model C\_df\_0.71\_inter and
the corresponding `plateau' model C\_df\_p\_0.71 (the latter having a
lower density in the inner region around the system) allows us to
evaluate the impact of the azimuthal dependence of the outflow
density, as the line of sight probes different parts of the outflow
spiral structure. As discussed in Sect.~\ref{sec:spiral}, the
'plateau' model mimicks situations corresponding to orbital phases
where the line of sight does not enter the cone of opening
$\theta_{\rm flow}$ (Fig.~\ref{fig:global}), thus when the density
close to the star is low. Figure~\ref{fig:SED_plateau}, which presents
the SED in the UV regime for these two models, reveals that the
'plateau' model has no far-UV absorption, whereas the non-plateau
models have.  Thus, one expects an orbital modulation of the far-UV
continuum as a consequence of the azimuthal distribution of the
outflowing material close to the star. A confrontation with far-UV
spectro-photometric observations could thus possibly provide insights
into the spatial structure of the outflow.

\subsubsection{Emission lines from weakly ionised species}
\label{sec:lines}

Figure~\ref{fig:SED} reveals the presence of many \ion{Fe}{ii}
emission lines for the different models considered, which are best
seen in the range 0.2 -- 0.3~{\textmu}m but are also present in the
optical. Figure~\ref{fig:emissivity} reveals that these lines mainly
form in the outer part of the outflow, outside the Roche
lobe. Detecting these lines may help to prove that some material is
indeed escaping from the binary system. The absence of such lines in
spectra of genuine Algols may thus be considered as an argument
against the presence of outflowing matter in these systems.

\subsubsection{Infrared} 
\label{sec:infrared}

\begin{figure}
  \centering \includegraphics[width=0.48\textwidth]{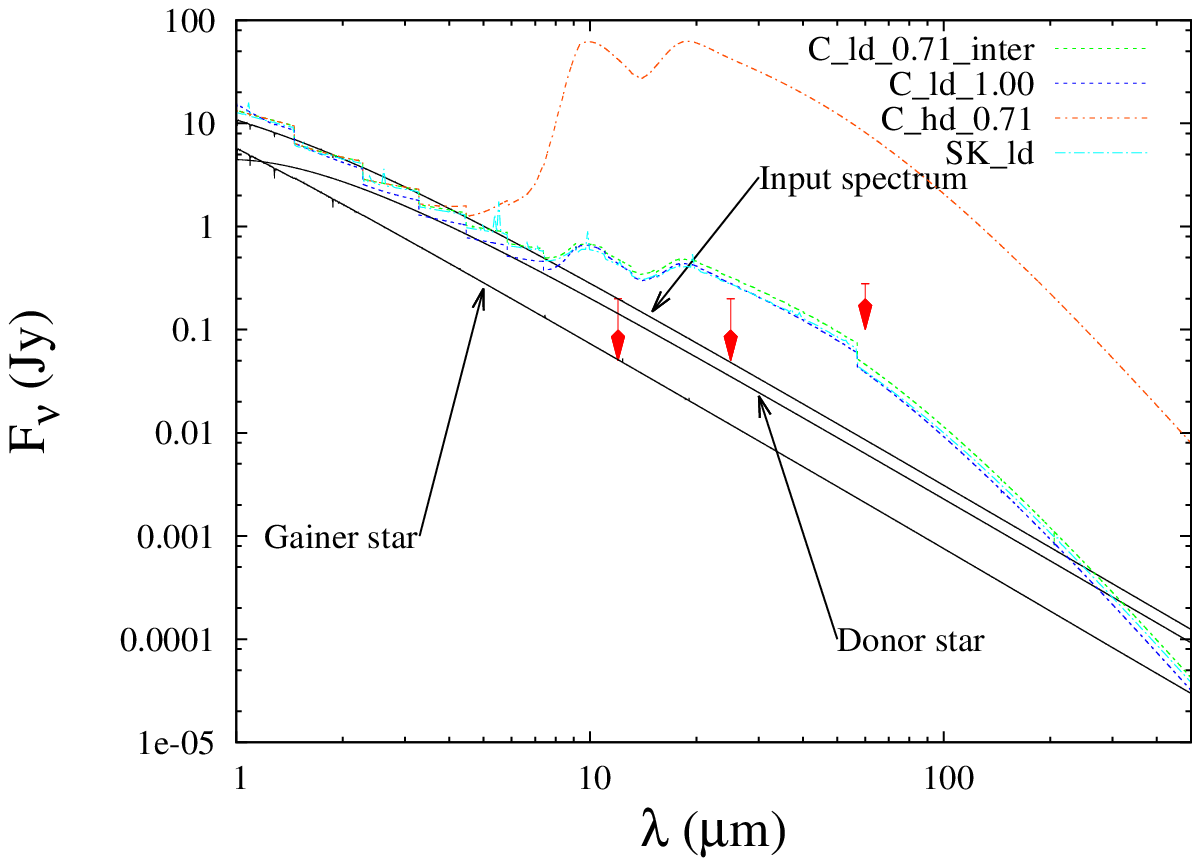}
  \caption{Same as Fig.~\ref{fig:SED} but for the IR region, including
    the \textsc{Skirt} model C\_SK\_ld. For clarity, spectral lines
    have been removed from the \textsc{Cloudy} SEDs. The red arrows
    are 5~$\sigma$ IRAS upper limits (in the faint-source catalogue),
    namely 0.2~Jy for the 12 and 25~{\textmu}m bands and 0.28~Jy for
    the 60~{\textmu}m band.}
  \label{fig:SED_IR}
\end{figure}

Figure~\ref{fig:SED_IR} shows the characteristic silicate feature that
is produced by the dust grains in the outer part of the outflow in our
models. All models produce infrared excesses above the 5$\sigma$ IRAS
detection limit at 12 and 25~{\textmu}m, but only the simulation with
the largest dust-to-gas ratio has a 60~{\textmu}m flux detectable by
IRAS (for systems located as usual at a distance of 300~pc).  In the
`plateau' simulation (C\_ld\_p\_0.71, not shown in the figure), there
is no dust emission, because the temperature falls below the
condensation temperature at a location where the density is too low to
produce any detectable infrared excesses.

\subsubsection{Synthetic photometry}
\label{sec:colors}

\begin{figure}
  \centering \includegraphics[width=0.48\textwidth]{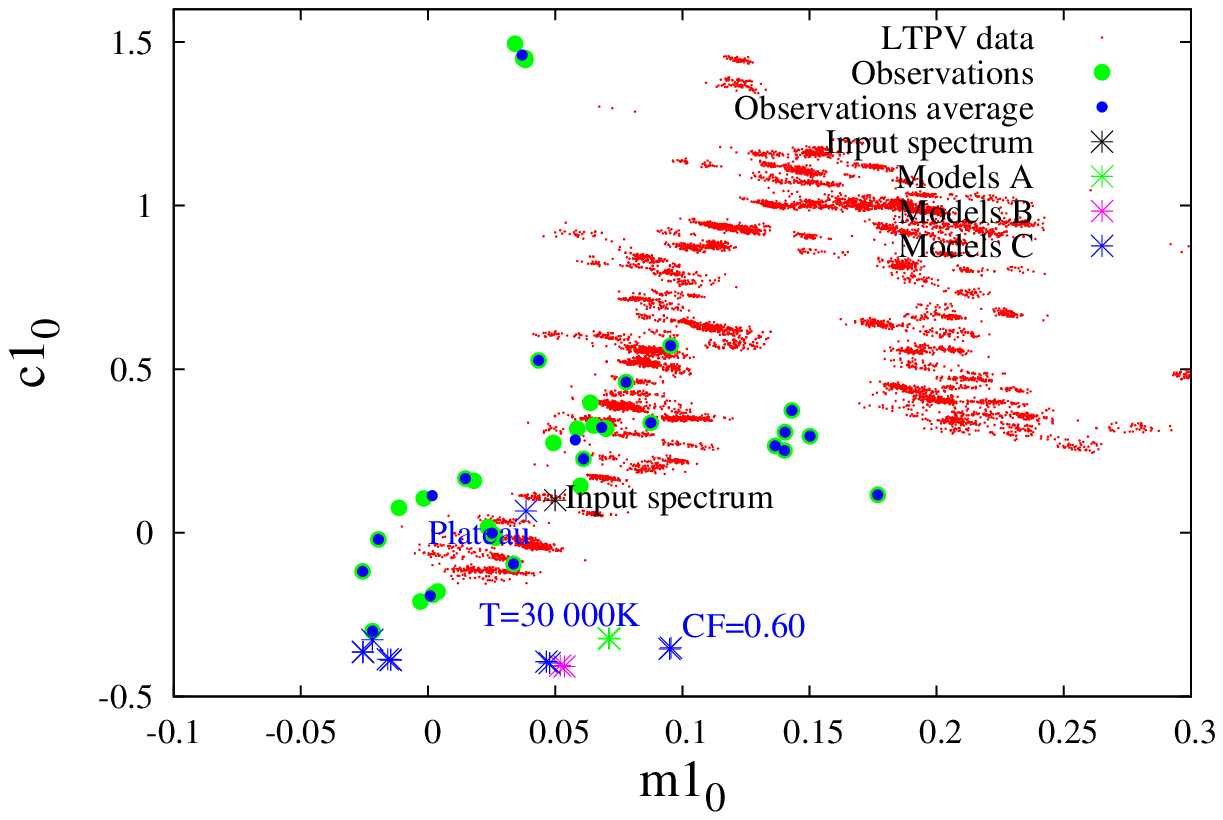}
  \caption{Str\"omgren photometry [$m1_0$ ; $c1_0$] diagram for our
    models (crosses) and for comparison normal main-sequence stars
    (red dots) from the Long-Term Photometry of Variables project
    \citep[][to be discussed in
      Sect.~\ref{sec:obs:stromgren}]{1991A&AS...87..481M}. From the
    latter, only data obtained with the Danish telescope (system~7)
    were retained. For clarity, the models are not labelled and only
    their general properties are indicated in blue. We refer to
    Table~\ref{tab:colorstrom} for the values. We added observations
    of W~Ser, B[e] and Be stars for comparison (green dots). The blue
    dots are the average values for the latter objects when several
    observations are available.}
  \label{fig:mix_photo}
\end{figure}

\begin{table}
  \caption{Magnitudes in Str\"{o}mgren ($uvby$) bands and
    corresponding colour indices.}  \centering
  \label{tab:colorstrom}
  \setlength{\tabcolsep}{3pt}
  \begin{tabular}{lccccccc}
    \hline
    \hline\\
    Model & $u$ & $v$ & $b$ & $y$ & $b-y$ & $m_1$ &
    $c_1$\\ \\
    \hline \\
    A\_df\_0.71&5.08&5.31&5.22&5.20&0.02&0.07&-0.32\\
    A\_ld\_0.71&5.09&5.31&5.22&5.20&0.02&0.07&-0.32\\
    B\_df\_0.71&5.33&5.64&5.54&5.49&0.05&0.05&-0.41\\
    B\_ld\_0.71&5.33&5.64&5.54&5.49&0.05&0.05&-0.41\\
    C\_df\_p\_0.71&5.75&5.68&5.67&5.71&-0.03&0.04&0.07\\
    C\_df\_1.00\_inter&5.21&5.48&5.42&5.34&0.08&-0.02&-0.33\\
    C\_ld\_1.00\_inter&5.21&5.48&5.42&5.34&0.08&-0.02&-0.33\\
    C\_df\_0.71\_inter&5.08&5.37&5.27&5.15&0.11&-0.02&-0.39\\
    C\_df\_0.71\_cold&5.17&5.48&5.39&5.35&0.04&0.05&-0.40\\
    C\_ld\_0.71\_inter&5.08&5.37&5.27&5.15&0.12&-0.01&-0.39\\
    C\_ld\_0.71\_cold&5.18&5.48&5.39&5.35&0.04&0.05&-0.39\\
    C\_hd\_0.71\_inter&5.08&5.37&5.27&5.16&0.12&-0.01&-0.39\\
    C\_ld\_0.71\_hot&5.07&5.34&5.25&5.14&0.12&-0.03&-0.36\\
    C\_df\_0.71\_hot&5.07&5.34&5.25&5.14&0.12&-0.03&-0.36\\
    C\_ld\_0.60\_inter&5.15&5.42&5.33&5.33&-0.01&0.10&-0.35\\
    C\_df\_0.60\_inter&5.15&5.42&5.33&5.33&-0.01&0.10&-0.35\\
    Stellar SED (C)&5.38&5.28&5.28&5.33&-0.05&0.05&0.10\\
    \hline
  \end{tabular}
  \tablefoot{$m_1=(v-b)-(b-y)$; $c_1=(u-v)-(v-b)$}
\end{table}

\begin{table}
  \caption{Magnitudes in SDSS bands and corresponding colour
    indices.}  \centering
  \label{tab:colorsdss}
  \setlength{\tabcolsep}{3pt}
  \begin{tabular}{lccccccc}
    \hline
    \hline\\
    Model & $u^\prime$ & $g^\prime$ & $r^\prime$ & $i^\prime$ & $z^\prime$ & $u^\prime-g^\prime$ &
    $g^\prime-r^\prime$ \\ \\
    \hline \\
    A\_df\_0.71&4.83&5.06&5.23&5.40&5.86&-0.22&-0.17\\
    A\_ld\_0.71&4.84&5.06&5.23&5.40&5.86&-0.22&-0.17\\
    B\_df\_0.71&5.09&5.36&5.51&5.68&6.15&-0.27&-0.15\\
    B\_ld\_0.71&5.09&5.36&5.51&5.68&6.15&-0.27&-0.15\\
    C\_df\_p\_0.71&5.43&5.51&5.81&6.04&6.23&-0.08&-0.30\\
    C\_df\_1.00\_inter&4.96&5.22&5.25&5.44&5.71&-0.26&-0.03\\
    C\_ld\_1.00\_inter&4.96&5.22&5.25&5.44&5.71&-0.26&-0.03\\
    C\_df\_0.71\_inter&4.84&5.08&5.24&5.44&5.89&-0.25&-0.16\\
    C\_df\_0.71\_cold&4.93&5.22&5.39&5.57&6.02&-0.28&-0.18\\
    C\_ld\_0.71\_inter&4.84&5.08&5.24&5.44&5.89&-0.25&-0.16\\
    C\_ld\_0.71\_cold&4.94&5.22&5.39&5.57&6.02&-0.28&-0.17\\
    C\_hd\_0.71\_inter&4.84&5.08&5.24&5.44&5.89&-0.25&-0.16\\
    C\_ld\_0.71\_hot&4.82&5.05&5.15&5.42&5.87&-0.23&-0.10\\
    C\_df\_0.71\_hot&4.82&5.05&5.15&5.42&5.87&-0.23&-0.10\\
    C\_ld\_0.60\_inter&4.90&5.17&5.36&5.55&5.99&-0.27&-0.19\\
    C\_df\_0.60\_inter&4.90&5.17&5.36&5.55&6.00&-0.27&-0.19\\
    Stellar SED (C)&5.30&5.33&5.68&5.97&6.21&-0.03&-0.35\\
    \hline
  \end{tabular}
\end{table}

\begin{table*}
  \caption{Magnitudes in 2MASS and WISE bands assuming the models are
    located 300~pc away from the sun.} \centering
  \label{tab:colorwise}
  \setlength{\tabcolsep}{3pt}
  \begin{tabular}{lccccccccccccc}
    \hline
    \hline\\
    Model & $J$ & $H$ & $K_{\mathrm{s}}$ & $W1$ & $W2$ & $W3$ &
    $W4$ &$F_J/F_{K_{\mathrm{s}}}$ & $F_{W4}/F_{W1}$& $F_{W3}/F_{W1}$\\ 
    \hline \\
    A\_df\_0.71&5.24&5.46&5.23&5.31&5.23&5.01&4.70&2.39&0.05&0.14\\
    A\_ld\_0.71&5.23&5.46&5.23&5.31&5.23&4.19&3.22&2.41&0.19&0.29\\
    B\_df\_0.71&5.52&5.72&5.45&5.46&5.27&4.95&4.58&2.27&0.06&0.16\\
    B\_ld\_0.71&5.52&5.72&5.44&5.46&5.28&4.24&3.35&2.26&0.19&0.31\\
    C\_df\_p\_0.71&5.67&5.60&5.51&5.43&5.35&4.93&4.58&2.09&0.06&0.16\\
    C\_df\_1.00\_inter&5.44&5.68&5.59&5.77&5.71&5.46&5.08&2.80&0.05&0.14\\
    C\_ld\_1.00\_inter&5.44&5.68&5.56&5.77&5.70&4.61&3.41&2.80&0.24&0.30\\
    C\_df\_0.71\_inter&5.39&5.63&5.44&5.51&5.41&5.11&4.71&2.54&0.06&0.15\\
    C\_df\_0.71\_cold&5.45&5.67&5.42&5.45&5.31&4.99&4.64&2.37&0.06&0.16\\
    C\_ld\_0.71\_inter&5.39&5.63&5.44&5.51&5.40&4.49&3.27&2.54&0.21&0.26\\
    C\_ld\_0.71\_cold&5.44&5.67&5.43&5.46&5.32&4.21&3.20&2.39&0.22&0.33\\
    C\_hd\_0.71\_inter&5.39&5.64&5.44&5.49&5.17&-0.21&-1.99&2.54&26.43&19.43\\
    C\_ld\_0.71\_hot&5.37&5.63&5.45&5.53&5.41&4.35&3.17&2.59&0.24&0.30\\
    C\_df\_0.71\_hot&5.37&5.63&5.45&5.53&5.42&5.14&4.74&2.60&0.06&0.15\\
    C\_ld\_0.60\_inter&5.38&5.59&5.31&5.34&5.21&4.20&3.12&2.27&0.21&0.29\\
    C\_df\_0.60\_inter&5.38&5.59&5.31&5.34&5.23&4.93&4.60&2.28&0.05&0.15\\
    Stellar SED (C)&5.76&5.77&5.76&5.76&5.75&5.73&5.74&2.41&0.04&0.11\\
    \hline
  \end{tabular}
  \tablefoot{$F_J/F_{K_{\mathrm{s}}}$, $F_{W4}/F_{W1}$ and
    $F_{W3}/F_{W1}$ are the flux ratios.}
\end{table*}

To ease the comparison of our models with observations,
Table~\ref{tab:colorstrom} presents, for all our models, the
magnitudes in the Str\"{o}mgren $uvby$ system and the corresponding
colour indices, using the passbands and zero points of
\cite{2011PASP..123.1442B} and assuming that the models are located
300~pc away from the sun. As expected, the systems with a strong
Balmer continuum emission appear with a negative value of
$c_1=(u-v)-(v-b)$.  Figure~\ref{fig:mix_photo} presents our models in
the [$m_{1,0}$ ; $c_{1,0}$] diagram compared with data from the
Long-Term Photometry of Variables project
\citep{1991A&AS...87..481M}. Only comparison stars (labelled A, B, or
C in their catalogue) have been retained to define the main
sequence. The $c_{1,0}$ and $m_{1,0}$ indices are not affected by
interstellar reddening and are defined by \citep{1966ARA&A...4..433S}:
\begin{align}
  c_{1,0} =& c_1 - 0.2(b-y) = \left[(u-v)-(v-b)\right] - 0.2(b-y)\\
  m_{1,0} =& m_1 + 0.18(b-y) = \left[(v-b)-(b-y)\right] + 0.18(b-y).
\end{align}
While our input spectrum falls within the main-sequence population
\citep{1966ARA&A...4..433S}, our models considerably deviate towards
lower $c_{1,0}$. Deviations in $m_{1,0}$ are also observed for some
models.  In addition to the Str\"{o}mgren photometry,
Table~\ref{tab:colorsdss} shows the SDSS magnitudes and two color
indices, which we calculated using the passbands provided by
\citet{2000A&AS..147..361M} and the zero points from
\citet{1996AJ....111.1748F}. We also present, in
Table~\ref{tab:colorwise}, the 2MASS and WISE band magnitudes that
will be used in Sect.~\ref{sec:wise}. WISE magnitudes are computed
according to \cite{2005PASP..117..421T}, based on transmission curves
from the Explanatory Supplement to the WISE All-Sky Data Release
Products
\citep{2012yCat.2311....0C}\footnote{\url{http://wise2.ipac.caltech.edu/docs/release/allsky/}
  \url{expsup/index.html}, 2014-02-27, Sect. IV.4.h, Figure 5a and
  Table 3}. 2MASS response curves are set up according to
ADPS\footnote{\url{http://ulisse.pd.astro.it/Astro/ADPS/Systems/}
  \url{Sys\_new\_021/index\_n021.html}, 2014-03-05} (See also the
``Explanatory Supplement to the 2MASS All Sky Data Release and
Extended Mission
Products''\footnote{\url{http://www.ipac.caltech.edu/2mass/releases/}
  \url{allsky/doc/explsup.html}, 2014-03-05, Sect. VI.4.a}).

\subsubsection{Radio emission: \ion{H}{i}-21~cm and rotational CO 
  lines}
\label{sec:radio}

The 21~cm line of neutral hydrogen (\ion{H}{i}-21~cm) and molecular
rotational lines such as CO$_{1-0}$~$\lambda=2.6806$~mm and
CO$_{2-1}$~$\lambda=1.3004$~mm may help to map the cold (neutral)
outer regions of the outflow. Although the 21~cm line may be
completely blended with ISM lines, the high speed of the ejected
material (with terminal velocities up to $\approx$\,900~km~s$^{-1}$)
may help to distinguish between the circumstellar and ISM
contributions. In our models, the CO abundance is very small and does
not lead to any observable lines. The \ion{H}{i}-21~cm line is also
very weak, possibly because of the low densities far from the star, in
the neutral hydrogen region. The \ion{H}{i}-21~cm line luminosity does
not exceed a few $10^{20}\ \mathrm{erg\,s}^{-1}$ for both the
dust-free and dusty simulations, corresponding to a few {\textmu}Jy at
a resolving power of 2\,000 and a distance of 300 pc. The outer part
of the outflow is therefore probably not detectable in these radio
bands.

\subsection{Models with strong ionisation}
\label{sec:hotspot_configuration}

\begin{figure}[t]
  \centering
  \includegraphics[width=0.49\textwidth]{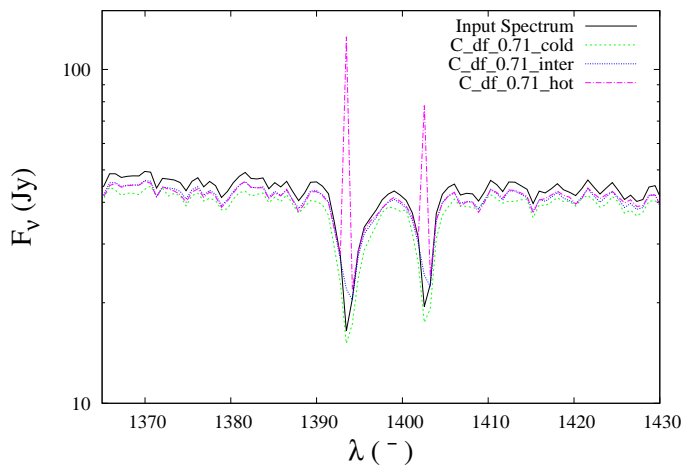}
  \caption{[\ion{Si}{iv}] 1394 and 1403$\AA$ doublet for three
    different configurations.}
  \label{fig:siIV}
\end{figure}

In this section, we investigate the impact of the hotspot temperature
and SED on the emergent SED as well as on the formation of dust and
emission lines.

A hotspot SED corresponding to a higher temperature implies a higher
number of ionising photons and thus larger abundances of species such as
 \ion{Si}{iv}, \ion{N}{v} or \ion{C}{iv}. Figure~\ref{fig:siIV}
shows the [\ion{Si}{iv}] doublet at 139.4~nm and 140.3~nm. There is no
difference between the incident and emerging SEDs for the
model with $T_\mathrm{hs}=30\,000$~K ('C\_df\_0.71\_low'). 
With a hotter hotspot and star, emission lines
from strongly ionised elements can form (although still only barely noticeable for
$T_\mathrm{hs}=35\,000$~K; model 'C\_df\_0.71\_inter').

The bottom panel of Fig.~\ref{fig:emissivity} shows the relative
abundance of \ion{Si}{iv} and \ion{Fe}{ii} for the models
C\_df\_0.71\_inter and C\_df\_0.71\_hot. We find that \ion{Si}{iv} is
restricted to the Roche volume around the gainer star.  We therefore
expect a strong line-profile variability due to the orbital motion, as
well as eclipses by the donor and the gainer stars as observed in some
W~Ser systems (see Appendix~\ref{ap:WSer_UV}). Unfortunately, we
cannot simulate these effects in our models, which do not resolve the
inner binary system. Due to the short cooling time-scale close to the
star (Fig.~\ref{fig:temp}), only the material above the hotspot
(within the cone of opening angle $\theta_\mathrm{flow}$ in
Fig.~\ref{fig:global}) may contain highly ionised species. The low
hotspot-temperature models produce much less \ion{Si}{iv}, but more
\ion{Fe}{ii} than the high-temperature ones.

The observed presence of [\ion{Si}{iv}] emission in the Algol-related
systems W~Ser (Sect.~\ref{sec:obs:NAlgols} and
Appendix~\ref{ap:WSer_UV}) does not necessarily imply systemic mass
loss however, because it probes the inner part of the nebula, which
may still be gravitationally bound. However, if present, the
variability of lines from such highly ionised species may help to
understand the ejection mechanism \citep{1989SSRv...50...35G}. Indeed,
since the ionised elements form close to the hot region (the ejection
point in the hotspot paradigm), we should be able to constrain the
position of the UV source and check whether it is compatible with the
location of the accretion point. On top of that, radial-velocity
measurements may constrain the global motion of this material
(expelled from the accretion region or revolving around the accreting
star, ...).

With increasing hotspot temperatures, only the outermost regions
contain dust. Thus the following dichotomy is expected: [dust/weak
  ionisation] or [no dust/strongly ionised elements], which is
supported by observations as discussed in Sect.~\ref{sec:obs:Algols}.

\subsection{Parameter study: geometry - \textsc{Skirt} simulations
  - dust-to-gas ratio}
\label{sec:geo}

\begin{figure}[t]
  \centering \includegraphics[width=0.49\textwidth]{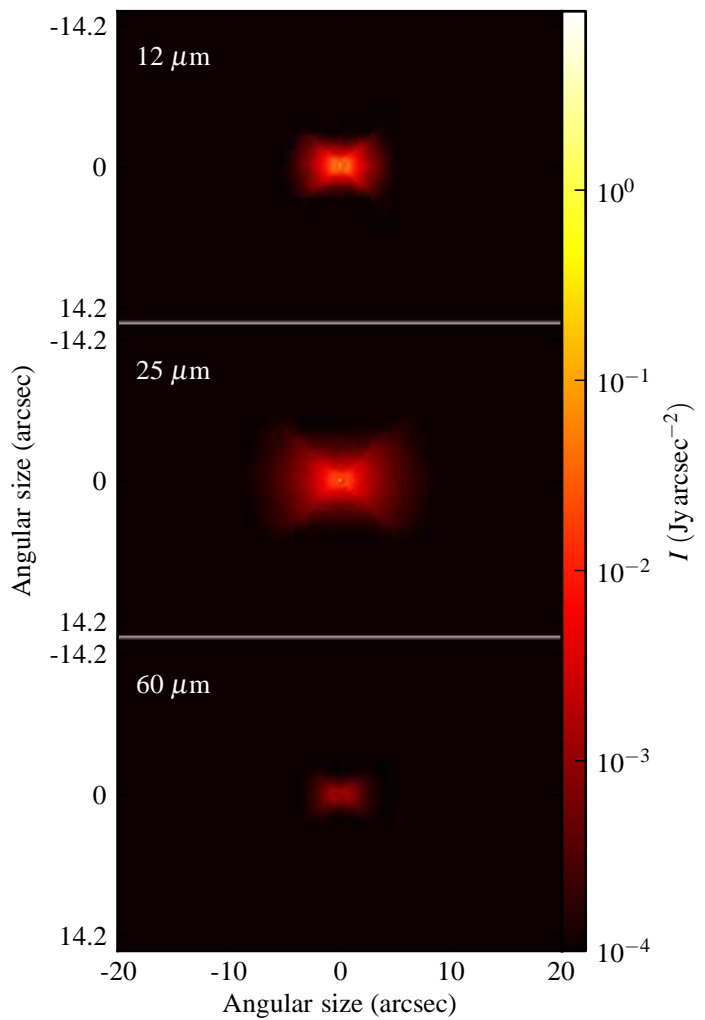}
  \caption{\textsc{Skirt} intensity map at 12~{\textmu}m (\emph{top}),
    25~{\textmu}m (\emph{middle}) and 60~{\textmu}m (\emph{bottom})
    for $\Phi = 2 \times 10^{-5}$ (model C\_SK\_ld). To ease the
    comparison, the same limits are used for the three bands for the
    `x' and `y' axes. The angular size and intensity are given for a
    system at 300~pc. The figure is zoomed on a box
    $12\,000\times8\,516$~AU$^2$ wide.}
  \label{fig:2D}
\end{figure}

In this section, we compare 1D results from \textsc{Cloudy} with
results from the 3D dust radiative-transfer code \textsc{Skirt}. When
the codes are applied to the same environment (as described by the
stellar parameters, the opening angle of the stream, ...), a very good
agreement between the 1D and 3D results is found (see
Fig.~\ref{fig:SED_IR}: compare the dashed green line and dotted
long-dashed cyan line). This agreement validates the results obtained
with \textsc{Cloudy} and described so far.

As described in Sect.~\ref{sec:codes}, we calculated models with
different covering factors $f_\mathrm{c}$ representing different
outflow geometries. Figure~\ref{fig:SED} shows the impact of the
covering factor for three configurations: the spherical emission case
($f_\mathrm{c}=1$) and two cases of a disc-like emission
($f_\mathrm{c}=0.71$ and $f_\mathrm{c}=0.6$). For each covering
factor, we recompute the density profile according to Eq.~\ref{eq:rho}
so that the total mass in the outflow is conserved. The UV/optical
absorption increases with decreasing covering factor, because the
densities increase, but the effect is modest. With the higher covering
factors, the IR flux is significantly reduced because the dust condensation
radius being larger, less dust
contributes to the infrared emission (Fig.~\ref{fig:SED_IR}).

Figure~\ref{fig:2D} shows a 2D map of the edge-on system at three
different wavelengths (12, 25, and 60~{\textmu}m) produced with
\textsc{Skirt} using the same physical properties as in model
C\_ld\_0.71\_inter (same opening angle and dust-to-gas ratio $\Phi = 2
\times 10^{-5}$; model C\_SK\_ld). We restrain our investigation to
regions where the flux is of the order of actual detection limits, set
at 0.1~mJy~arcsec$^{-2}$ ($\approx$\,1/10 of current observational
thresholds in the mid- to near-IR for the VLT/VISIR instrument;
\citealt{2004Msngr.117...12L}). We do not expect a strong contribution
from the cosmic infrared background (CIB) at the selected
wavelengths. \citet{1998ApJ...500..525S} show that at 100~{\textmu}m
(their lowest wavelength), the CIB value is only
$\approx$\,3.3~mJy~arcsec$^{-2}$ in the galactic plane and much less
closer to the galactic pole and at shorter wavelengths. The speckle
effect (coarseness observed especially far from the centre of the box)
seen in Fig.~\ref{fig:2D} is due to the smaller number of photon
packages sent from the outer regions compared to the inner regions; it
is thus numerical noise. The edges of the disc-like outflow are too
cold to be seen at 12~{\textmu}m and only the longer wavelengths can
probe these outer regions. However, the flux decreases at longer
wavelengths and drops below the adopted detection limit. The `X' shape
is due to the opening angle of the disc-like outflow, seen through the
optically thin outer part. We remind that, in our models, there is no
material outside the opening angle $\theta_{\mathrm{flow}}$, whereas
in a real system we expect a much smoother transition, with the
highest density on the binary plane and a lower density at higher
latitudes. Therefore, the `X' will be less marked.  The inner hot
region, where dust cannot form, has a diameter of $\approx 425$~AU
corresponding to $\approx 1\farcs{4}$ (18~pixels) in
Fig.~\ref{fig:2D}. For comparison, the binary system has a separation
of less than 1~AU and is thus far from being resolved.

For the models (C\_hd\_0.71 and C\_SK\_hd) with a higher dust-to-gas
ratio ($\Phi = 4 \times 10^{-3}$, typical of the ISM) and a covering
factor $f_\mathrm{c}=0.71$, the 12~{\textmu}m peak rises up to
$\approx60$~Jy in both the \textsc{Cloudy} and the \textsc{Skirt}
simulations, as compared to 0.6~Jy for the model C\_ld\_0.71\_inter
displayed in Figs.~\ref{fig:SED_IR} and \ref{fig:2D}. A ratio of 100
between the 12~{\textmu}m fluxes of models differing by a factor of
200 in their dust-to-gas ratios may seem surprising when the dust
shell is optically thin, as is the case here. However, one should
first subtract the stellar contribution from the observed
12~{\textmu}m flux of model C\_ld\_0.71\_inter, and then take the
ratio with the flux from model C\_hd\_0.71. The expected ratio of 200
is found then. Clearly, the uncertainty on the dust-to-gas ratio in
the Algol environment has a strong impact on the present predictions.

\section{Confrontation to observations}
\label{sec:obs}

\subsection{Stellar samples}
\label{sec:obs:NAlgols}

In this section, we present the samples of objects investigated for
the presence of systemic mass loss as predicted by the models. The
observable diagnostics extracted from the models and discussed in
Sect.~\ref{sec:results} are not typical of Algol systems, but share
similarities with the properties of Algol-related systems like
W~Serpentis and $\beta$~Lyr\ae{} stars, B[e] and Be stars, Symbiotic
Algols and Double Periodic Variables.

\paragraph{Algols.} The definition of this class adopted by the Simbad
database\footnote{\url{http://simbad.u-strasbg.fr/simbad/} is
  ``detached eclipsing binaries showing the Algol paradox (the more
  massive component of the system is the less evolved)''. However,
  many among the systems considered as Algols in the literature are
  actually semi-detached systems.} In the present study, we call
``genuine Algols'' systems that are detached or semi-detached, showing
the Algol paradox in the absence of UV emission lines (that are used
as a property of W~Ser systems as discussed next). We created our own
sample of Algols based on the catalogues of \cite{1980AcA....30..501B}
containing 701
Algols\footnote{\url{http://cdsarc.u-strasbg.fr/viz-bin/Cat?II/150A}}
and \cite[][]{2004A&A...417..263B} containing 435
Algols\footnote{\url{http://cdsarc.u-strasbg.fr/viz-bin/Cat?J/A\%2bA/417/263}}. In
these two catalogues, the classification `Algol' is based on the
individual mass ratio.

\paragraph{W~Serpentis and $\beta$~Lyr\ae{}} are the prototypes of two
subsets of Class-II\footnote{rapid mass transfer systems with mass
  ratio reversed \citep{2013A&A...557A..40D}} Algol systems. The
accreting star in these objects may be obscured by a thick accretion
disc (see the review of \citealt{1989SSRv...50...35G}). They
correspond to rapid mass-transfer systems most likely during case B,
or BB for $\beta$~Lyr itself (second phase of mass transfer following
the case-B mass-transfer phase). The main property of W~Ser systems is
the presence of strong ultraviolet emission lines ([\ion{C}{iv}],
[\ion{C}{ii}], [\ion{He}{ii}], [\ion{N}{v}], [\ion{Si}{iv}]...; see
Appendix~\ref{ap:WSer_UV}) seen at every phase
\citep{1980HiA.....5..831P, 1982NASCP2338..526P, 1988ESASP.281a.221P,
  1989SSRv...50...95P, 1989SSRv...50...35G}. Our sample of W Ser is
built on the catalogue of 31 peculiar emission-line Algols (PELAs;
\citealt{1996ASPC...93..312G}) after removing from that list those
systems with no evidence for the presence of these UV emission
lines. The strong ultraviolet emission lines from highly ionised
species observed in W~Ser systems may be identified with those
appearing in our model with the highest-temperature hotspot
(C\_df\_0.71\_hot). However these spectral signatures cannot confirm
the presence of material escaping the system since they form inside
the Roche lobe of the star (see
Sect.~\ref{sec:hotspot_configuration}). Moreover, in W~Ser systems,
there are two contributions to the strongly ionised emission lines
(\citealt{1995ApJ...447..401W}, see Appendix~\ref{ap:WSer_UV}): the
boundary layer between the disc and the star, responsible for the
broad emission seen at every phase and the hotspot, responsible for
the highly phase-dependent narrow emission. There have also been
reports of infrared emission that we will further investigate.
\cite{1997AstL...23..698T} report infrared excesses at 3.5 and
5~{\textmu}m possibly attributable to dusty material. Similarly,
RY~Sct possesses an extended dusty nebula (2\,000~AU wide;
\citealt{2007ApJ...667..505G}).

\paragraph{Be stars} are rapidly rotating (v>300 km\,s$^{-1}$),
non-supergiant B stars showing strong Balmer emission lines or
evidence of an ionised shell (see \citealt{2013A&ARv..21...69R} for a
recent review). These stars form a dust-free, outwardly diffusing,
disc. In single Be stars, the formation mechanism of these decretion
discs is related to the rapid rotation of the star, but some Be stars
are interacting binaries, likely Algols. In this case, the formation
of the disc and the star's rapid rotation could result from the RLOF
accretion stream \citep{1991ApJ...370..597P, 1991MNRAS.253...55C,
  2013A&A...557A..40D}. Although there is no catalogue of truly
“Algol-like” Be stars yet available, \citet{2001PAICz..89....9H} has
compiled an extensive list of Be stars in binary systems that may
serve as an interesting starting point. In that list, we identified
five Algol-like systems of particular interest as it will be apparent
from our subsequent analysis: \object{$\phi$ Per}
\citep{1998ApJ...493..440G}, \object{SX Aur}
\citep{1988ApJ...327..265L}, HR~2142
\citep{1983PASP...95..311P,1991A&A...250..437W}, $\pi$~Aqr
\citep{2002ApJ...573..812B} and CX~Dra \citep{2002A&A...394..181B}.

\paragraph{B[e] stars} are B stars showing strong forbidden emission
lines, Balmer emission lines and infrared excesses due to hot
circumstellar dust \citep{1998A&A...340..117L}. However, this class
contains several poorly understood systems and includes young and
evolved, single and binary stars. Therefore, different mechanisms at
the origin of the gas surrounding B[e] stars may be involved, among
which some may be related to mass transfer across the B[e] binary
systems \citep[see, e.\,g.,][]{2012A&A...542A..50D}. There is no
specific catalogue of binary B[e] stars.

\paragraph{Symbiotic Algols.} Although most symbiotic stars are
assumed to host an accreting white dwarf
\citep[e.\,g.,][]{2007BaltA..16....1M}, it is not necessarily the case
for all of them. For example, SS~Lep \citep{2011A&A...536A..55B} is a
non-eclipsing symbiotic, showing the Algol paradox, composed of a
red-giant donor and a main-sequence accreting star. This system shows
the presence of a dusty circumbinary disc or envelope
\citep{2001ApJ...550L..71J}. The main difference with standard Algols
lies in the mass transfer scheme, which is believed to be Wind
Roche-lobe overflow (WRLOF; \citealt{2007ASPC..372..397M}) instead of
standard RLOF. Due to the long period of the system (260\fd{3}), mass
accretion onto the gainer is supposed to occur via an accretion
disc. Unfortunately, the properties of the hotspot on the edge of an
accretion disc are badly constrained, which makes it difficult to
predict systemic mass-loss rates and to compare with observational
data. Moreover, it is not clear whether the hotspot mechanism can work
with WRLOF, because of the lower mass-transfer rates.

\paragraph{Double periodic variables} (DPVs) are interacting binaries
exhibiting two closely linked periodic variations, the shortest one
being the orbital period
\citep{2003A&A...399L..47M,2005MNRAS.357.1219M, 2008MNRAS.389.1605M},
and the second one of unknown nature but probably linked to the
variations in the strength of the disc wind
\citep{2012MNRAS.427..607M}. They are considered as one specific
evolutionary step for the more massive Algols. An interesting property
of these objects is the surprising constancy of their orbital period,
which is not expected in Algols undergoing RLOF mass transfer with a
mass ratio different from 1 \citep{2013MNRAS.428.1594G}. An
explanation could be that systemic mass loss compensates for the
period increase caused by mass transfer from a less massive star to
its more massive companion. \cite{2008MNRAS.389.1605M} already
emphasised the need for further investigation of these systems.  Some
DPVs (the Galactic ones) already appear in our Algol or W~Ser
catalogues but most of these objects have been found thanks to surveys
in the Small Magellanic Cloud and are therefore too far away to allow
any clear detection of an infrared excess. We therefore do not include
them in this study.

\subsection{Constraint on the dust formation: IR
  diagnostics}
\label{sec:obs:Algols}

In this section, we search for systems that exhibit infrared colours
typical of dust emission. We select Class~II and III objects (flaring
and quiescent Algols, respectively, with a mass ratio already
reversed, according to the nomenclature of Paper~I), for which we
expect most of the mass to have been ejected, and dust to have
formed. Class~III objects should no longer show any signature of
systemic mass loss, though, because of the short-living outflow, but
they were nevertheless considered for completeness. We extracted the
IR data from IRAS, 2MASS and WISE catalogues. The use of the IRAS
catalogues is motivated by their completeness in the 12 to
60~{\textmu}m range. On the other hand, the choice of the WISE and
2MASS databases has been motivated by their better detection limits as
compared to IRAS (IRAS sensitivities: 0.5~Jy at 12, 25 and
60~{\textmu}m; WISE 5$\sigma$ sensitivities: 120, 160, 650 and
2600~{\textmu}Jy at 3.3, 4.7, 12 and 23~{\textmu}m, respectively;
2MASS 10$\sigma$ at $J<15.0$~mag, $H<14.2$~mag,
$K_{\mathrm{s}}<13.5$~mag).

\subsubsection{IRAS data}
\label{sec:iras}

We selected the data from the IRAS (InfraRed All-sky Survey)
point\footnote{\url{http://cdsarc.u-strasbg.fr/viz-bin/Cat?II/125}}
\citep{1988iras....7.....H} and
faint\footnote{\url{http://cdsarc.u-strasbg.fr/viz-bin/Cat?II/156A}}
\citep{1990IRASF.C......0M} source catalogues. We retrieve 241 Algol
entries with a matching IRAS source within a 2{\arcmin} positional
tolerance. We search for IR excesses using the colour indices
\begin{equation}
  [i]-[j] = -2.5 \log \left( \frac{F_{i}}{F_{j}} \frac{F_{j,0}}{F_{i,0}}\right),
\end{equation}
where $i$ and $j$ correspond to the IRAS bands at 12, 25, or
60~{\textmu}m, and $F_{j,0}$ and $F_{i,0}$ are the zero-magnitude
fluxes. Rejecting spurious detections (contaminated by a nearby
infrared source), we did not find any systems presenting a deviation
larger than 0.5~mag above the black-body line in the
([12]-[25],[12]-[60]) plane.

\subsubsection{WISE data}
\label{sec:wise}

\begin{figure*}[t]
  \centering
  \includegraphics[width=1.0\textwidth]{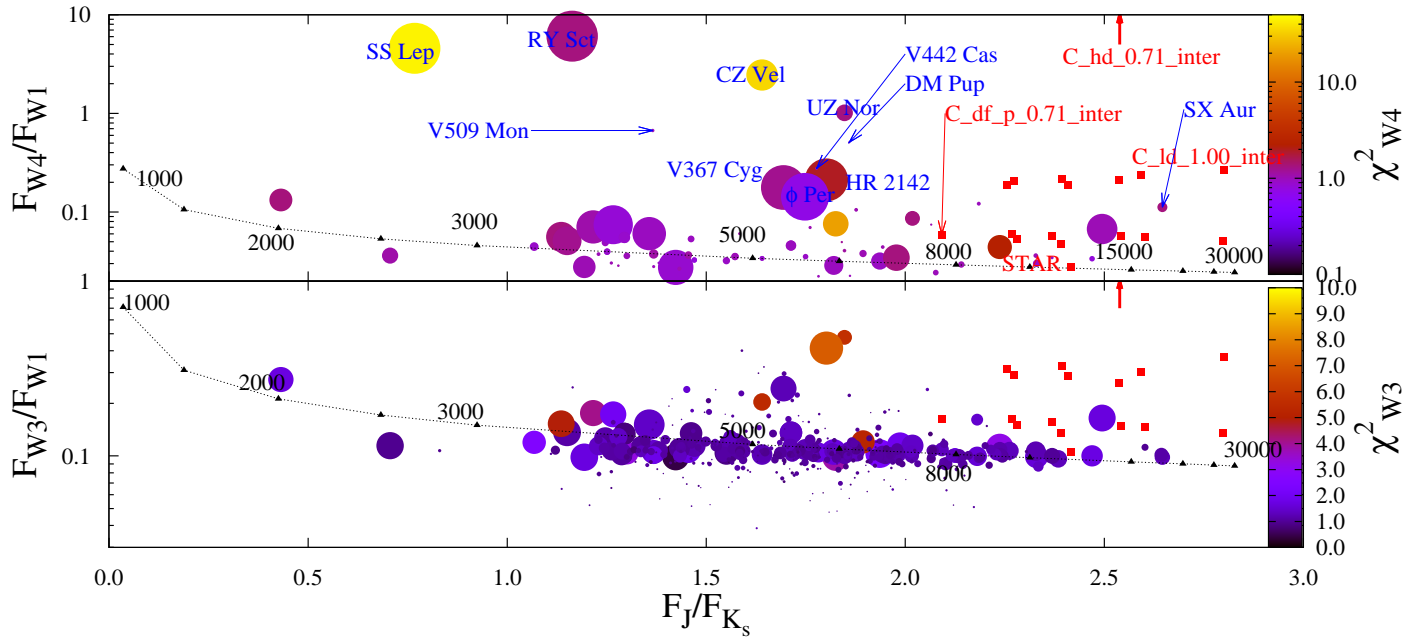}
  \caption{\textbf{Top panel:} WISE~$F_{W4}/F_{W1}$ against
    2MASS~J/K$_{\mathrm{s}}$ flux ratios for Algols from the catalogue
    of \cite{2004A&A...417..263B}, and for W Ser from the list of
    \cite{1996ASPC...93..312G}. The symbiotic Algol SS~Lep and a few
    Be and B[e] systems have been considered as well (as discussed in
    Sect.~\ref{sec:obs:NAlgols}). The black filled triangles (linked
    by the dashed line) are black bodies of given temperatures (labels
    in Kelvin). The red filled squares represent the simulations
    (STAR: input spectrum). For clarity, only extreme models are
    labelled, we refer to Table~\ref{tab:colorwise} for unlabelled
    models. The filled circles correspond to Algols with a
    SNR~\textgreater~3.2. The size of the circle is proportional to
    the SNR for band $W4$ (except for SS~Lep which has a SNR of 962
    and for which we limit the size of the symbol) and the symbol
    colour relates to the $\chi^{2}$ for the same band. \textbf{Bottom
      panel:} Same as top panel but for $F_{W3}/F_{W1}$ with SNR and
    $\chi^{2}$ relative to $W3$.}
  \label{fig:wise}
\end{figure*}

\begin{table*}
  \caption{2MASS and WISE magnitudes for the 13 systems deviating the
    most from the $F_{W4}/F_{W1}$ black-body law (see
    Fig.~\ref{fig:wise}).}
  \label{tab:wise}\setlength{\tabcolsep}{3pt}
  \centering
  \begin{tabular}{lrc@{}c@{}cccccc}
    \hline
    \hline\\
    \textbf{Name} & $P_{\mathrm{orb}}$ & (Sp.1) & $J$ & $H$ & $K_{\mathrm{s}}$ & $W1$
    & $W2$  & $W3$  & $W4$ \\
    & (d) &  & (mag.) & (mag.) & (mag.) & (mag.) & (mag.) & (mag.) &
    (mag.)\\
         & $q$ & [Sp.2] &$\varpi$ (mas)&  &  & [SNR$_{\mathrm{W1}}$
    ]($\chi^{2}_{\mathrm{W1}}$) & 
    [SNR$_{\mathrm{W2}}$]($\chi^{2}_{\mathrm{W2}}$) &
    [SNR$_{\mathrm{W3}}$]($\chi^{2}_{\mathrm{W3}}$) &
    [SNR$_{\mathrm{W4}}$]($\chi^{2}_{\mathrm{W4}}$)\\ \\ 
    \hline\\
    \noalign{\textbf{Algols}}
    
    \textbf{CZ Vel} & 5.1927$^{(1)}$ & (B3)$^{(3)}$ & 9.441 & 9.192 &9.032&
    8.562 & 8.485 & 7.814 & 3.669\\
          &0.50$^{(3)}$&[B6]&Unk.&&&[50](1.75)&[52.6](1.79)&
    [48.2](5.46)&[42.9](35.4)\\
   
    \textbf{UZ Nor} & 3.1960$^{(2)}$ & (A1)$^{(2)}$ & 10.262 & 10.079 & 9.982 &
    9.873 & 9.817 & 8.202 & 5.937\\
          &0.18$^{(3)}$&[K0IV]&Unk.&&&[48.3](3.27)&[57.4](2.82)&
    [40.8](5.7)&[22.9](1.33)\\
 
    \textbf{V509 Mon} & 4.9174$^{(3)}$ & (B7)$^{(3)}$ & 12.595 & 12.171
    & 11.987 & 11.868 & 11.905 & 11.515 &8.38\\
         &0.32$^{(3)}$&[G4IV]&Unk.&&&[46.3](1.23)&[43.9](1.37)&
    [6.5](1.14)&[4.7](0.992)\\

    \textbf{DM Pup} & 3.567$^{(13)}$ & (B5)$^{(3)}$ & 12.761 & 12.569
    & 12.488 & 12.057 & 12.035 & 11.215 & 8.868\\
        &Unk.&[A2.5]&Unk.&&&[46.1](1.38)&[42.6](1.61)&
    [9.6](1.00)&[3.2](1.09)\\

    \textbf{V442 Cas} & 3.5922$^{(3)}$ & (A7)$^{(3)}$ & 11.729 & 11.516
    & 11.407 & 11.286 & 11.271 & 10.961 &8.76\\
        &0.18$^{(3)}$&[K2IV]&Unk.&&&[49.6](1.36)&[52.2](1.42)&
    [14.9](0.995)&[4.2](0.994)\\
      
    \textbf{DH Her} & Unk. & (A5)$^{(6)}$ & 10.054 & 9.997
    & 9.956 & 10.134 & 10.168 & 10.189 & 8.449\\
         &Unk.&[Unk.]&Unk.&&&[47.9](5.90)&[53.6](6.06)&
    [26.9](1.12)&[5.4](0.979)\\\\
    
    \noalign{\textbf{$\beta$ Lyr}}
    
    \textbf{RY Sct} & 11.12$^{(9)}$ & (O9.7Ibpe)$^{(10)}$ & 6.245 &
    5.854 & 5.463 & 4.955 & 3.895 & 0.014 & -0.922\\
          &0.23$^{(11)}$&[O6.5 I]&1.59&&&[14.2](12.1)&[19.4](51.0)&
    [21.3](0.550)&[83.96](2.21)\\\\

    \noalign{\textbf{W Ser}}

    \textbf{V367 Cyg} & 18.6$^{(14)}$ & (A7I)$^{(3)}$ & 5.202 & 5.006
    & 4.828 & 4.72 & 4.038 & 3.781 & 2.676 \\
            &0.825$^{(14)}$&[Unk.]&1.74&&&[13.2](0.434)&[20.9](2.34)&
    [70.9](1.27)&[31.4](1.24)\\ 

    \textbf{W Ser} & 14.154$^{(3)}$ & (B emb.)$^{(12)}$ & 7.416 & 6.109
    & 5.558 & 5.106 & 4.560 & 4.038 & 3.379 \\
        &0.64$^{(3)}$&[F5III]&0.70&&&[13.6](0.737)&[18.2](3.65)&
    [69](1.72)&[31.3](1.36)\\\\

    \noalign{\textbf{Symbiotic Algol}}

    \textbf{SS Lep} & 260.3$^{(5)}$ & (A1V)$^{(5)}$ & 2.094 & 2.105 & 1.671 &
    3.455 & 1.536 & -1.307 & -2.117\\
           &0.47$^{(5)}$&[M6II]&3.59&&&[39](186)&[64.3](195)&
    [3](0.0094)&[962](45.2)\\\\

    \noalign{\textbf{Be}}

    \textbf{HR 2142} & $\sim$80$^{(17)}$ & (B2IVne)$^{(17)}$ & 5.088 & 5.068
    & 4.781 & 4.828 & 3.756 & 3.31 &2.592\\
       & Unk. &[Unk.]&2.48&&&[12.1](15.4)&[16](25.5)&
    [91.1](7.12)&[58.9](1.96)\\
    
   \textbf{$\pi$ Aqr} & 84.1$^{(16)}$ & (B1Ve)$^{(4)}$ & 5.305 & 5.365 & 5.351 &
    4.573 & 3.767 & 4.056 & 3.58\\
       &0.16$^{(16)}$&[A-F]$^{(16)}$&4.17&&&[13.8](1.22)&[16](2.10)&
    [73.8](1.68)&[41.7](1.02)\\

    \textbf{CX Dra} & 6.696$^{(18)}$ & (B2.5Ve)$^{(18)}$ & 5.870 & 5.805 & 5.486 &
    5.457 & 5.140 & 5.128 & 4.854\\
       &0.23$^{(16)}$&[F5III]$^{(18)}$&2.52&&&[17.4](1.27)&[32.9](2.07)&
    [77.9](4.48)&[49](1.11)\\\\

    \textbf{$\phi$ Per} & 127$^{(15)}$ & (B2e)$^{(15)}$ & 4.049 & 3.955
    & 3.709 & 3.68 & 2.774 & 2.639 & 1.869 \\
        &0.158$^{(3)}$&[sdO]&4.54&&&[8.7](0.247)&[12.2](1.59)&
    [53.5](2.03)&[65.4](0.737)\\ 

    \textbf{SX Aur} & 1.21$^{(4)}$ & (B2)$^{(4)}$ & 8.431 & 8.471 & 8.541 &
    8.594 & 8.655 & 8.636 & 7.047\\
        &0.61$^{(8)}$&[B5]&1.39&&&[47.9](2.97)&[55.6](3.53)&
    [42](1.29)&[13.6](1.34)\\

    \hline
  \end{tabular}
  \tablefoot{`Unk.' refers to unknown values, `emb.' to an embedded
    star where the spectral type is only roughly guessed. $q$ is the
    mass ratio defined as $q=M_{\mathrm{donor}}/M_{\mathrm{gainer}}$.}
  \tablebib{ $^{(1)}$ \citealt{2009NewA...14..129Z}; $^{(2)}$
    \citealt{2011NewA...16..157Z}; $^{(3)}$
    \citealt{2004A&A...417..263B} and references therein; $^{(4)}$
    \citealt{1988ApJ...327..265L}; $^{(5)}$
    \citealt{2011A&A...536A..55B}; $^{(6)}$
    \citealt{2004AcA....54..207K}; $^{(7)}$
    \citealt{1985JBAA...96...27I}; $^{(8)}$
    \citealt{1988ApJ...327..265L}; $^{(9)}$
    \citealt{2004AcA....54..207K}; $^{(10)}$
    \citealt{1982AJ.....87.1300W}; $^{(11)}$
    \citealt{2007ApJ...667..505G}; $^{(12)}$
    \citealt{1989SSRv...50...95P}; $^{(13)}$
    \citealt{1988BICDS..35...15H}; $^{(14)}$
    \citealt{2001A&A...368..932Z}; $^{(15)}$
    \citealt{2001A&A...368..471H}; $^{(16)}$
    \citealt{2002ApJ...573..812B}; $^{(17)}$
    \citealt{1991A&A...250..437W}; $^{(18)}$
    \citealt{2002A&A...394..181B}.}
\end{table*}

2MASS\footnote{\url{http://cdsarc.u-strasbg.fr/viz-bin/Cat?II/246}}
and WISE\footnote{\url{http://cdsarc.u-strasbg.fr/viz-bin/Cat?II/311}}
data are provided by \citet{2003yCat.2246....0C} and
\citet{2012yCat.2311....0C}, respectively. Figure~\ref{fig:wise} shows
the flux ratios [WISE~$F_{W4}$ (22.1~{\textmu}m)/$F_{W1}$
  (3.35~{\textmu}m)] and [WISE~$F_{W3}$ (11.6~{\textmu}m)/$F_{W1}$
  (3.35~{\textmu}m)] as functions of [2MASS~$F_{J}$
  (1.25~{\textmu}m)/$F_{K_{\mathrm{s}}}$ (2.17~{\textmu}m)] for 79
Algols having positional matches with both 2MASS and WISE within less
than 3{\arcsec} (31 objects have matches better than 0\farcs12 and 15
better than 0\farcs06). Spurious detections due to a diffraction spike
from a close bright star, the persistence from a bright star scanned
immediately before the target, halo scattering from a close source or
optical ghost have all been discarded. We retained systems having a
signal-to-noise ratio (SNR) in band $W4$ higher than 3.2. In addition,
we note that the variability labels range from `not variable' (about
45 objects; this number depends on the band considered) to `high
probability of being true variable' (34 objects). This variability
observed by WISE could be attributed to an outflow, but it could also
simply result from the eclipse of the donor by the gainer (in our
model, the donor is three times brighter than the gainer in the $W4$
band). A more precise analysis of variability with WISE data is beyond
the scope of this study.

Figure~\ref{fig:wise} includes observed systems (filled circles) and
our models (red squares). The circle colour codes the $\chi^{2}$
(which measures the deviation from the point-spread function--PSF; it
is larger than 1 for extended sources) for bands $W4$ and $W3$. The
circle size is proportional to the SNR in bands $W3$ and $W4$. The $J$
and $K_{\mathrm{s}}$ bands are not affected by dust emission (see
Fig.~\ref{fig:SED_IR}), and only depend on the stellar temperatures
and on the covering factor. Since our simulations are based on one
specific Algol system at three different stages of its evolution (see
details in Table~\ref{tab:char}), the range in
$F_{J}/F_{K_{\mathrm{s}}}$ values covered by the models is limited
compared to observations. In the adopted Algol model system, the donor
star dominates the spectrum at longer wavelengths, beyond the $J$ band
(Figs.~\ref{fig:SED_IR} and \ref{fig:2D}).

On the other hand, $F_{W4}/F_{W1}$ directly traces the presence of IR
emission from dust grains (Fig.~\ref{fig:SED_IR}), and
$F_{W3}/F_{W1}$ probes the silicate emission around 12~{\textmu}m.
The models distribute along three horizontal sequences in the upper
panel of Fig.~\ref{fig:wise}, dust-free models corresponding to the
lowest $F_{W4}/F_{W1}$ value (close to the stellar locus), followed by
low-dust models at an intermediate $F_{W4}/F_{W1}$ value, and finally
the high-dust model 'C\_hd\_0.71\_inter' falling above the limit of
the figure.

Fig.~\ref{fig:wise} reveals that several Algols and a few W~Ser and
$\beta$~Lyr in our sample have $F_{W4}/F_{W1}$ ratios falling above
the black-body sequence, thus hinting at the presence of dust
emission. The same holds true for the $F_{W4}/F_{W2}$ ratio, which has
not been displayed though because the $W2$ flux may reach (close to)
saturation for bright systems. For four Algols (\object{CZ Vel},
\object{UZ Nor}, \object{DM Pup} and \object{V509 Mon}), one
$\beta$~Lyr (\object{RY Sct}), and the symbiotic Algol \object{SS
  Lep}, $F_{W4}/F_{W1}$ is close to or larger than 1, well above our
model predictions for a dust-to-gas ratio of
$\Phi=2\times10^{-5}$. The SNRs of the WISE measurements for these
systems are among the highest in our sample (see
Table~\ref{tab:wise}). The WISE and 2MASS data for these
systems\footnote{We do not confirm thus the conclusion of
  \citet{1997AstL...23..698T} who seemingly found evidence for dust
  around the W~Ser system RX~Cas, though at shorter wavelengths (in
  the $LM$ bands at 3 -- 5 {\textmu}m). For this system, we found no
  significant infrared excesses from WISE
  ($F_{W4}/F_{W1}=0.056$). Therefore, the excess found by
  \citet{1997AstL...23..698T} in the $LM$ bands cannot come from
  dust.} and others with moderate excesses are collected in
Table~\ref{tab:wise}. Among these, not much is known (apart from their
eclipsing light curve) for DM~Pup, V509~Mon, CZ~Vel, UZ~Nor, V442~Cas,
and DH~Her, but given the interesting nature of all the other stars
picked up by the $F_{W4}/F_{W1}$ criterion, a dedicated study of this
small list of not well-studied stars will certainly pay off. RY~Sct is
a very interesting contact binary consisting of two seemingly O-type
supergiants (but the 11~d orbital period makes it unlikely that the
components are really as extended as supergiants are), with dust
forming in an ionising radiation field, and with a radiation-driven
nebular wind with a kinematic age of 120~yr
\citep[e.g.,][]{2001AJ....122.2700S}.

\begin{figure*}[t]
  \centering
  \includegraphics[width=0.45\textwidth, bb=-12 -5 552 400, clip=true]{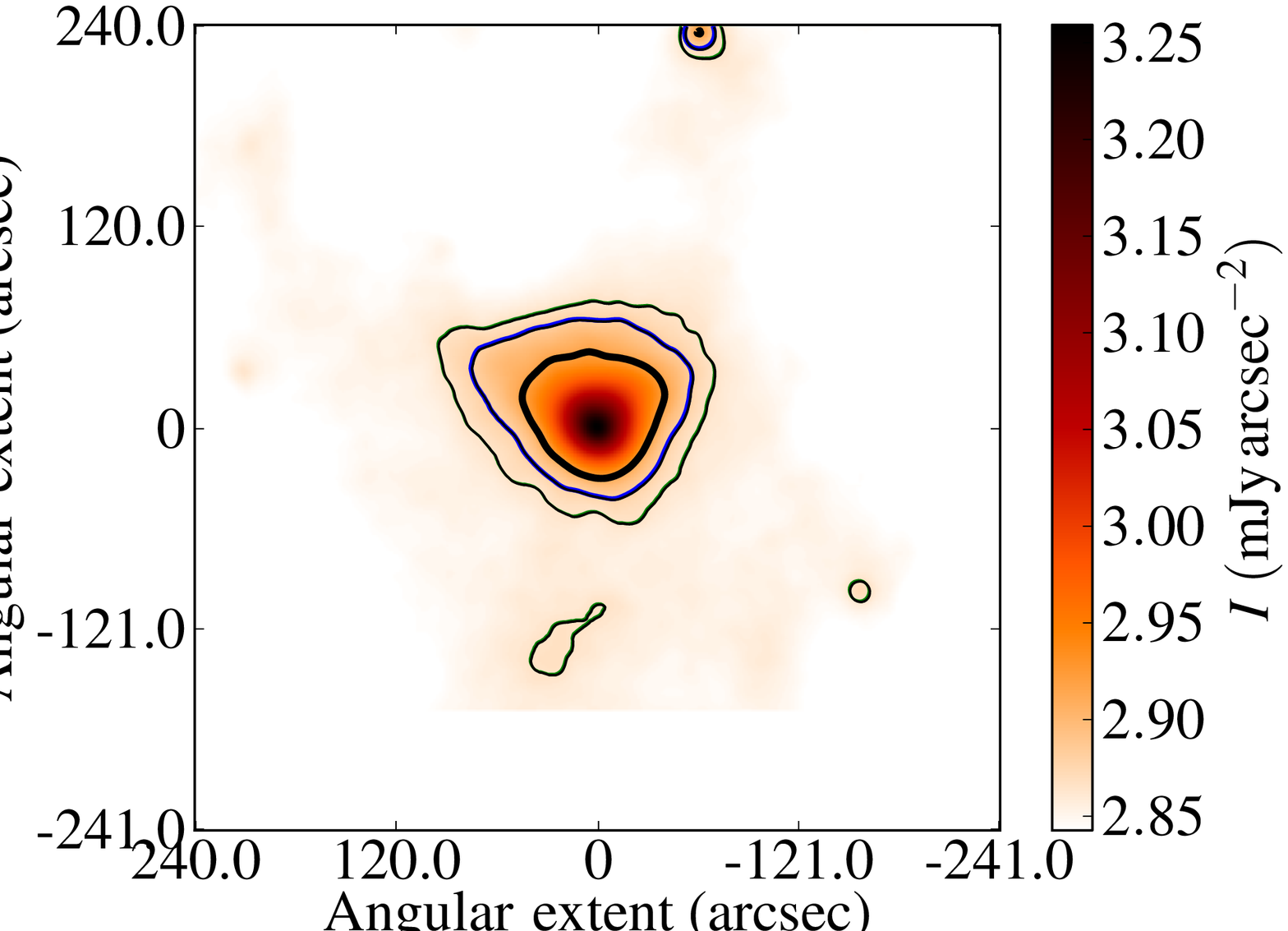}
  \includegraphics[width=0.45\textwidth, bb=-12 -5 552 400, clip=true]{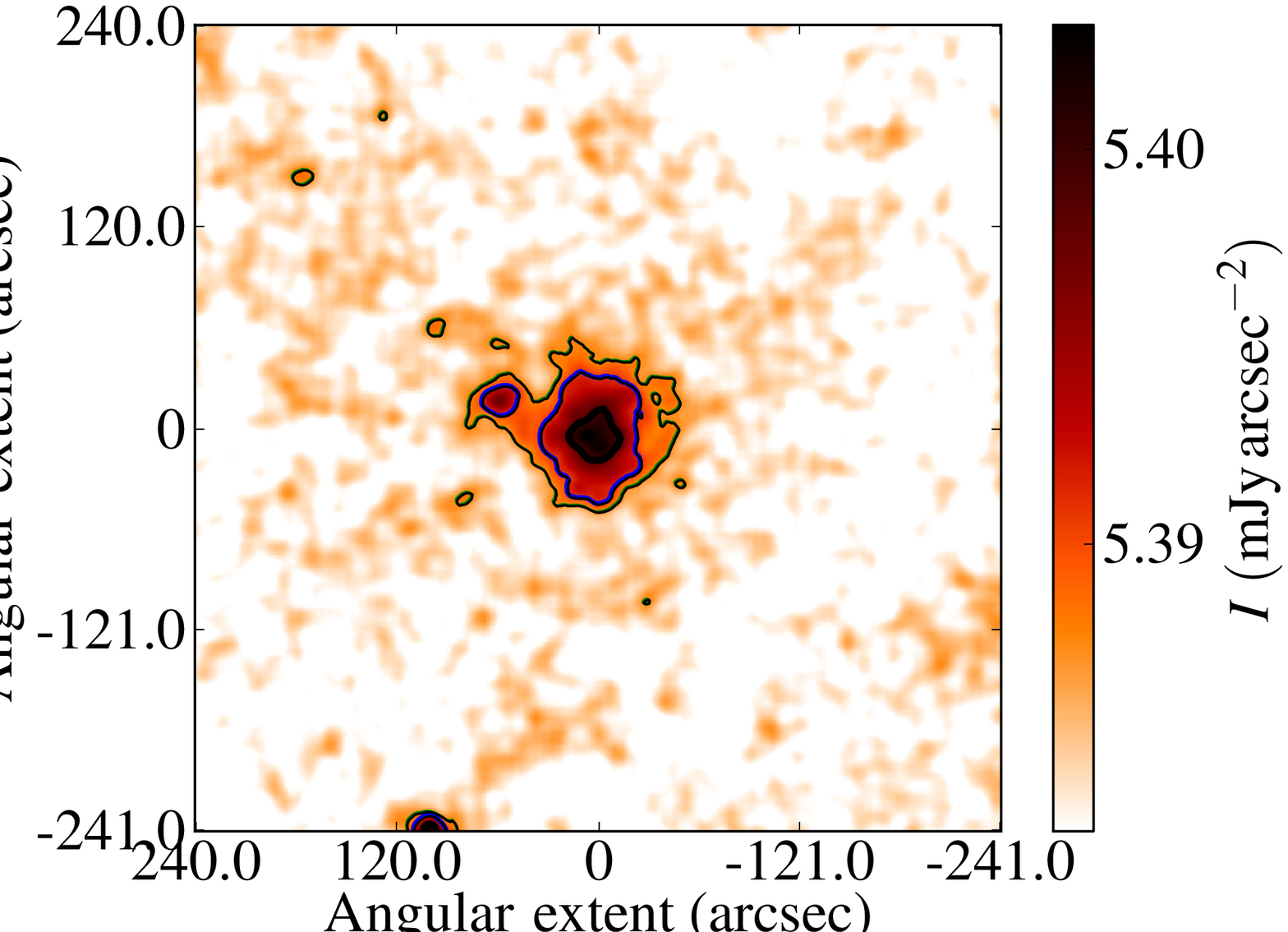}
  \caption{WISE band-4 images for the genuine Algol CZ~Vel (left) and
    the Be star SX~Aur (right).  The images are centred on the
    objects, with North up and East left. The flux scale goes from the
    background level to background plus 150 times the background
    scatter $\sigma$, and the first thin (green) contour line
    corresponds to the background level plus 3~times its fluctuation,
    the next intermediate (blue) contour curve to 5~$\sigma$ and the
    inermost thick black contour line to 10~$\sigma$.}
  \label{fig:imagewise}
\end{figure*}

Finally, three more systems have $F_{W4}/F_{W1}$ ratios close to the
value expected for $\Phi=2\times10^{-5}$: the Be systems \object{V696
  Mon} (=~HR~2142), \object{$\phi$ Per} and \object{SX Aur}. SX~Aur
shows an extended shell in the W4 band as can be seen in
Fig.~\ref{fig:imagewise}. This is an interesting result in view of the
controversy regarding the exact evolutionary status of that system:
from the light curve analysis, \cite{1979ApJ...228..828C} obtained a
semi-detached configuration with the secondary filling its Roche lobe,
and endorse an earlier suggestion by \cite{1973PASP...85..363S} that
the system is undergoing case A mass transfer and that mass reversal
has occurred. On the contrary, based on a new light curve analysis,
Linnell et al. (1988) concluded that the system is made of two
unevolved B stars, the primary being however on the verge of filling
its Roche lobe. Our discovery of an extended infrared shell around
SX~Aur (Fig.~\ref{fig:imagewise}) seems to suggest that some mass was
lost from the system in a recent past, even though mass transfer is
not currently occurring. This result confirms the interest of
considering Be stars as well when looking for evidence of systemic
mass loss from Algols (see Sect.~\ref{sec:obs:NAlgols}). Concerning
HR~2142, it must be noted that the target lies close to a star forming
region (Monoceros R2 at 56{\arcmin}) and it is likely that the
infrared emission does not originate from the binary system itself but
rather from the rich sky background.

Among the systems with the highest $F_{W4}/F_{W1}$ ratios, we note
that CZ~Vel is the only genuine Algol system with an extended nebula
detected by WISE (Fig.~\ref{fig:imagewise}), as could be inferred by
its high $\chi_{W4}^{2}$. Its largest angular size is about
180{\arcsec}. Unfortunately, the parallax of CZ~Vel is unknown, as is
therefore the linear size of the nebula. The nebula is very asymmetric
and does not resemble our predictions (Fig.~\ref{fig:2D}).

There is something else to remark from Fig.~\ref{fig:wise}. Systems
like W~Ser or V930~Oph (having the lowest $F_{J}/F_{K_{\mathrm{s}}} =
0.4$ and 0.71, respectively) exhibit very low equivalent black-body
temperatures ($T<3000$~K), in fact well below the components'
effective temperatures. These low temperatures thus indicate that the
system light is, at those infrared wavelengths, dominated by a very
cool object, possibly an accretion disc which obscures the gainer
star, as in W~Ser.  The very short period
\citep[$P=1\fd{4}$;][]{1993BICDS..42...27M} of \object{V930 Oph} does
not support, however, the presence of an accretion disc; direct impact
of the accreting stream is expected instead
\citep{1975ApJ...198..383L}. The low-temperature black body detected
is therefore likely to originate from a substantial amount of cool
dust far from the system, indicative of systemic mass loss.

Figure~\ref{fig:wise} shows that a typical ISM dust-to-gas ratio
($\Phi=4\times10^{-3}$, model C\_hd\_0.71) produces $F_{W4}/F_{W1}$
ratios larger than those observed with WISE in Algols with an infrared
excess. Thus, observations favour a smaller dust-to-gas ratio in the
stellar nebula than in the ISM. This conclusion is consistent with
dust being destroyed by high-energy photons from both the B star and
the hotspot. It is likely for the same reason that so few genuine
Algols exhibit infrared excesses caused by dust. Moreover, in the
framework of the hotspot mechanism (Paper~I), systemic mass loss
occurs in the early stage of mass transfer. Since the circumstellar
shell dissipates very quickly (within 300~years) after the cessation
of the non-conservative evolution, this is another reason why infrared
excesses are expected to be rare around genuine Algols.

Similarly, W~Ser systems rarely show the signature of dust
emission. This is compatible with our finding that the presence of
highly ionised species (as seen in W~Ser systems) is exclusive of the
presence of dust (see Sect.~\ref{sec:hotspot_configuration}).

\subsection{Constraint on Str\"{o}mgren photometry and weakly ionised
  emission lines}
\label{sec:obs:stromgren}

As mentioned in Sect.~\ref{sec:spectra_UV}, the presence of a strong
Balmer continuum in emission and the forest of metallic-line emission
in the emergent spectrum are not typical of Algols. Str\"{o}mgren
photometry, through its $c_1$ index (Sect.~\ref{sec:colors} and
Fig.~\ref{fig:mix_photo}), offers an easy way to check for the
presence of Balmer emission continuum. Usually, as concluded before,
IR excesses and emission lines or continuum are mutually
exclusive. There are three interesting exceptions, though: the
$\beta$~Lyr system RY~Sct, consisting of two stars with O-type spectra
\citep{2001AJ....122.2700S} and dust formation in the circumstellar
nebula despite ionising radiation \citep{1995ApJ...439..417G}. We
confirm the latter authors' finding, since RY~Sct is the system with
the largest $F_{W4}/F_{W1}$ ratio (Table~\ref{tab:wise}) and at the
same time the smallest $c_{1,0}$ index (Table~\ref{tab:stromgrenc1}),
indicative of a strong Balmer continuum in emission. The system
clearly has all features indicative of systematic mass loss. The Be
stars $\phi$~Per and $\pi$~Aqr also present a negative $c_{1,0}$,
indicative of a strong Balmer continuum. This property thus reveals
once more the interest of Be binary systems in the search for
Algol-like systems with systemic mass loss.
  
In Table~\ref{tab:stromgrenc1}, there are 3 other objects with
negative $c_{1,0}$ values, namely the W~Ser stars AX~Mon and V644~Mon,
and the Be star HR~2142. All the other systems are located in a region
of the $m_{1,0}$ -- $c_{1,0}$ diagram not typical of Balmer-continuum
emission. On the contrary, the W~Ser star V367~Cyg has a large
$c_{1,0}$ and deviates considerably from the main sequence
locus. \cite{1982ApJ...262..269Y} presented a dedicated study off
$uvby$ photometry of V367~Cyg (along with a few other W~Ser systems)
and found the same peculiar Str\"omgren colours as those reported
here, which imply either excess flux longward of 400~nm or a strong
absorption in the $u$ band, due for instance to improperly corrected
interstellar absorption. The latter interpretation is made likely by
the fact that V367~Cyg is a remote object in the Cygnus dust
clouds. Therefore, \cite{1982ApJ...262..269Y} conclude that ``there is
reason to question the reddening correction'' in such a peculiar
environment. Of course, one cannot firmly exclude an alternative
reason related to the physics of the binary system, despite the fact
that the ($m_{1,0}$, $c_{1,0}$) colours of V367~Cyg are not typical of
the other W~Ser systems studied here and by
\cite{1982ApJ...262..269Y}.

\begin{table}
  \caption{Str\"{o}mgren $m_{1,0}$ and $c_{1,0}$ for our sample of
    W~Ser and Be stars.}
  \label{tab:stromgrenc1}
  \begin{tabular}{lcccc}
    \hline
    \hline\\
    \textbf{Name} & Type & $m_{1,0}$ (mag.)& $c_{1,0}$ (mag.) & Ref. \\ \\
    \hline\\
    \textbf{RY Sct}  &$\beta$ Lyr& -0.0218 & -0.3002 & (4)\\
    \textbf{$\beta$ Lyr} &$\beta$ Lyr&0.0697&0.32& (12)\\
    &&0.07006&0.3186& (2)\\
    &&0.06518&0.3278& (6)\\
    \textbf{AU Mon}&W Ser&0.05866&0.3186& (1) \\
    &&0.04928&0.2748& (5) \\
    &&0.0599&0.144& (10)\\
    &&0.06386&0.3976& (7)\\
    \textbf{KX And} &W Ser&0.01472&0.1662& (7)\\
    \textbf{RZ Sct} &W Ser&-0.00166&0.1054& (1)\\
    &&-0.01142&0.0768& (9)\\
    &&0.01804&0.1594& (6)\\
    \textbf{UX Mon} &W Ser&0.14302&0.3742& (10)\\
    \textbf{V356 Sgr}&W Ser&0.0435&0.527& (9)\\
    \textbf{V367 Cyg} &W Ser&0.03818&1.4508& (1)\\
    &&0.03744&1.4494& (7) \\
    &&0.0383&1.445& (8)\\
    &&0.03424&1.4944& (6)\\
    \textbf{W Cru} &W Ser&0.41384&0.4614& (1)\\
    &&0.41672&0.4352& (14)\\
    &&0.4062&0.538& (9) \\
    \textbf{W Ser} &W Ser&0.14036&0.3076& (9)\\
    \textbf{V395 Aur} &Be&0.15024&0.2954& (6)\\
    \textbf{V617 Aur} &Be&0.14018&0.2518& (7) \\
    \textbf{V644 Mon} &Be&-0.01952&-0.0202& (10)\\
    \textbf{V1914 Cyg} &Be&0.17682&0.1162& (13)\\
    \textbf{U CrB} &Be&0.07786&0.4606& (6)\\
    \textbf{AX Mon} &Be&-0.02574&-0.1174& (6)\\
    \textbf{$\phi$ Per} &Be&0.0336&-0.095& (4) \\
    \textbf{CX Dra} &Be&0.06112&0.2262& (15) \\
    \textbf{$\pi$ Aqr} &Be&0.00188&-0.1892& (1)\\
    &&-0.00296&-0.2096& (2) \\
    &&0.0038&-0.179& (3)\\
    \textbf{HR 2142} &Be&0.02558&-0.0042& (1)\\
    &&0.0236&0.017& (2)\\
    &&0.02648&-0.0152& (3)\\
    \hline
  \end{tabular}
  \tablebib{ $^{(1)}$ \citealt{1998A&AS..129..431H};
 $^{(2)}$ \citealt{1976A&AS...25..213G};
 $^{(3)}$ \citealt{1971AJ.....76.1058C};
 $^{(4)}$ \citealt{1982A&AS...49..561W};
 $^{(5)}$ \citealt{1992A&AS...92..841K};
 $^{(6)}$ \citealt{1966GeoOM..21....0C};
 $^{(7)}$ \citealt{1983A&AS...54...55O};
 $^{(8)}$ \citealt{1969AJ.....74..705P};
 $^{(9)}$ \citealt{1983ApJS...52..429W};
 $^{(10)}$ \citealt{1985SAAOC...9...55K};
 $^{(11)}$ \citealt{1987A&AS...71..421R};
 $^{(12)}$ \citealt{1980A&AS...40....1H};
 $^{(13)}$ \citealt{1993A&AS..102...89O};
 $^{(14)}$ \citealt{1991A&AS...87..541G};
 $^{(15)}$ \citealt{1971AJ.....76.1058C}.}
\end{table}

\section{Conclusions}
\label{sec:conclusion}

We performed photoionisation and radiative-transfer simulations of the
material surrounding an Algol system that is loosing mass via the
hotspot mechanism. We calculated models with various configurations
(different hotspot temperatures, with or without dust, different
opening angles for the disc-like outflow), and find that the resulting
IR excesses produced by dust in the outer parts of the outflow are
consistent with WISE observations for a small fraction of Algols,
especially the systems CZ~Vel, UZ~Nor and V509~Mon. We also report
infrared excess in the Be system SX~Aur, in the $\beta$~Lyr\ae{}
system RY~Sct and in the symbiotic Algol SS~Lep (already known for the
latter two). As expected, most genuine Algols do not exhibit strong
infrared excesses since the systemic mass loss occurred at an earlier
phase and the material has since dissipated. Our models also predict
strong Balmer continua that are not observed in Algols but are visible
in some Be systems.

The outflowing gas can significantly reduce the observed UV
continuum-flux level. We suspect this absorption to quickly disappear
after the end of the non-conservative phase due to the high outflow
speed of the line-driven wind and to show some phase dependency as the
matter is launched from the hotspot. This variability may help to
identify systems undergoing systemic mass loss due to the hotspot
mechanism.

Our simulations indicate that the geometry of the outflow has an
impact on the emergent SEDs, especially in the IR range for dusty
models, where the fluxes decrease with increasing opening angle of the
outflow.

In the SEDs computed from our models with a hotspot temperature
$T_\mathrm{hs}\ga35\,000$~K, we find strong emission lines of highly
ionised species such as \ion{Si}{iv} which arise close to the stellar
surface, inside the Roche lobe of the gainer star, and are typical of
W~Ser systems. We emphasise the need for further observations of the
extended family of Algols, which includes DPVs, symbiotic Algols, Be
and B[e] binaries, W~Ser and $\beta$~Lyr systems. These types of
objects can be related to different steps in the evolution of Algols
or to different initial configurations (masses, periods). The search
for signatures of a non-conservative evolution should not focus
specifically on Algols, since they are older Class~III systems
(Paper~I), but more generally on Algol-related systems where systemic
mass loss can potentially occur. We also hope that this first study of
the hotspot-outflow formation will motivate a more precise analysis of
the stream/star or stream/disc interaction, preferentially by use of
3D models including radiation-hydrodynamics (see
\citealt{2000A&A...353.1009B,2013ARep...57..294N}, for $\beta$~Lyr).

\begin{acknowledgements}
  R.D.\ acknowledges support from the Communaut\'{e} fran\c{c}aise de
  Belgique -- Actions de Recherche Concert\'{e}es and benefits from a
  European Southern Observatory studentship. K.B.\ is an F.R.S.-FNRS
  post-doctoral fellow and acknowledges support from contract
  2.4501.12. L.S.\ is an FNRS Researcher. P.C.\ is supported by the
  CHARM network, part of the phase VII Interuniversity Attraction Pole
  (IAP) programme organised by the Belgian Science Policy Office
  (BELSPO). This research has made use of the VizieR catalogue access
  tool and SIMBAD database, CDS, Strasbourg, France. This publication
  makes use of data products from the Wide-field Infrared Survey
  Explorer, which is a joint project of the University of California,
  Los Angeles, and the Jet Propulsion Laboratory/California Institute
  of Technology, funded by the National Aeronautics and Space
  Administration, as well as of data products from the Two Micron All
  Sky Survey, which is a joint project of the University of
  Massachusetts and the Infrared Processing and Analysis Center,
  funded by the National Aeronautics and Space Administration and the
  National Science Foundation.
\end{acknowledgements}

\begin{appendix}
  \section{Estimation of the opening angle of the outflow}
  \label{ap:opening_angle}
  
  \begin{figure}
  \includegraphics[width=0.48\textwidth]{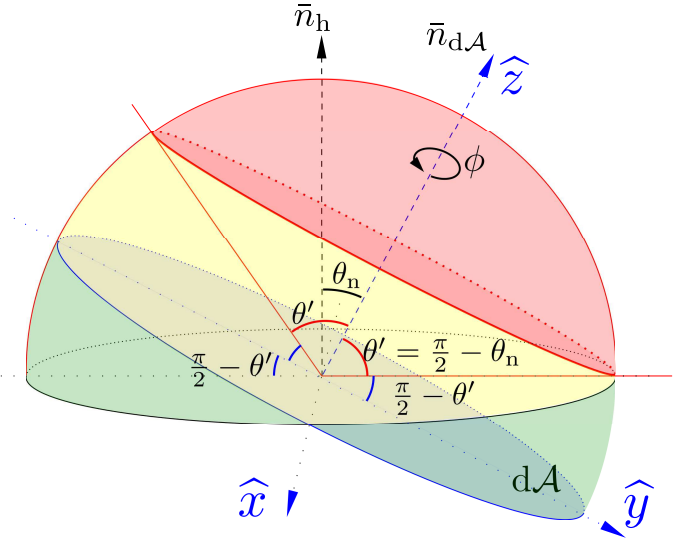}
  \caption{Geometry of the hotspot
    emission. $\bar{\vec{n}}_{\mathrm{h}}$ is the normal to the
    surface of the hotspot and $\bar{\vec{n}}_{\mathrm{d}\mathcal{A}}$
    is the normal of the surface $\mathrm{d}\mathcal{A}$, which forms
    an angle $\theta_{\mathrm{n}}$ with
    $\bar{\vec{n}}_{\mathrm{h}}$. $\theta_{\mathrm{flow}}/2$ (see
    text) is equal to $\theta_{\mathrm{n}}$ when the momentum flux
    integrated in azimuthal angle between $0$ and
    $\theta_{\mathrm{n}}$ equals 68\% of the momentum flux integrated
    over the total hemisphere (between 0 and $\frac{\pi}{2}$).}
  \label{fig:pressure_geometry}
\end{figure}

The geometry of the escaping stream may be complex, forming a disc-like
geometry, and possibly creating a spiral-like structure close to the
star (see Sect.~\ref{sec:spiral}). The geometrical thickness of the
disc structure is determined by the opening angle of the
outflow. Since the outflow is driven by the radiative flux of momentum
$\vec{p}_\nu$, we can estimate the geometrical thickness of the
outflow by use of the dependence of $\vec{p}_\nu$ on the direction
given by $\theta_n$ (angle measured with respect to the direction
perpendicular to the hotspot surface; see
Fig.~\ref{fig:pressure_geometry}).

The total momentum flux, normal to an element of surface
d$\mathcal{A}$ that is parallel to the hotspot, is given by
(Fig.~\ref{fig:pressure_geometry})
\begin{equation}
  \label{eq:momentum_flux}
  \vec{p}_{\nu}(\theta_\mathrm{n}=0) = \frac{I_{\nu}}{c} \bar{\vec{n}}_\mathrm{h}
         \int_{\mathrm{hemisphere}} \cos^{2}(\theta)\,\mathrm{d}\Omega ,
\end{equation}
where $I_{\nu}$ is the specific intensity (assumed to be isotropic)
originating from the hotspot, $\bar{\vec{n}}_\mathrm{h}$ is the unit
vector normal to the hotspot, $c$ is the speed of light and
$\mathrm{d}\Omega = \sin\theta\;\mathrm{d}\theta\;\mathrm{d}\phi$.

To calculate the momentum flux toward an arbitrary angle
$\theta_{\mathrm{n}}<\pi/2$, we introduce the reference frame
$(\widehat{x},\widehat{y},\widehat{z})$, where the $\widehat{x}$ axis
lies in the plane of the hotspot, the $\widehat{y}$ axis lies in the
plane $\mathrm{d}\mathcal{A}$, and $\widehat{z}$ is collinear with
$\bar{\vec{n}}_{\mathrm{d}\mathcal{A}}$, the normal to
$\mathrm{d}\mathcal{A}$. Note that $\theta$ is measured with respect
to $\bar{\vec{n}}_{\mathrm{d}\mathcal{A}}$. The general expression of
Eq.~\ref{eq:momentum_flux} for arbitrary angles $\theta_\mathrm{n}$
reads
\begin{equation}
  \label{eq:momentum_flux_general}
  \vec{p}_\nu(\theta_\mathrm{n})=\frac{I_{\nu}}{c}\bar{\vec{n}}_{\mathrm{d}
    \mathcal{A}}\int_\mathrm{hemisphere}\cos{\theta}\;\lvert\cos{\theta}
  \rvert\;\Theta(\bar{\vec{n}}\cdot\bar{\vec{n}}_\mathrm{h})\;\mathrm{d}\Omega,
\end{equation}
where $\bar{\vec{n}}$ is the unit vector pointing in the direction
$(\theta,\phi)$. The absolute value in the second factor of the
integrand ensures that radiation traversing the surface
$\mathrm{d}\mathcal{A}$ from above adds a negative contribution to the
total momentum flux. The Heavyside step function
$\Theta(\bar{\vec{n}}\cdot\bar{\vec{n}}_\mathrm{h})$ accounts for the
fact that the hotspot is only radiating upwards.

For a fixed angle $\theta_\mathrm{n}$, the vector
$\bar{\vec{n}}_\mathrm{h}$ normal to the hotspot has coordinates
$(0,-\sin\theta_\mathrm{n},\cos{\theta_\mathrm{n}})$, so that an
evaluation of the dot product gives the limits of integration in
$\phi$:
\begin{align}
  \phi_{\mathrm{lim,1}}& =
  \begin{cases}
    \arcsin{[1/(\tan{\theta}\tan{\theta_\mathrm{n}})]} \\
    0
  \end{cases}&
  \begin{array}{l}
    \theta>\pi/2-\theta_\mathrm{n} \\
    \mathrm{otherwise}
  \end{array}, \\
  \phi_{\mathrm{lim,2}}& =
  \begin{cases}
    \pi-\arcsin{[1/(\tan{\theta}\tan{\theta_\mathrm{n}})]} \\
    2\pi
  \end{cases}&
  \begin{array}{l}
     \theta>\pi/2-\theta_\mathrm{n} \\
     \mathrm{otherwise}
  \end{array},
\end{align}
where the latter expression for $\phi_{\mathrm{lim,2}}$ results from
the symmetry of the problem (the angle $\phi$ is 0 above the
$\widehat{x}$ axis).  With these limits, Eq.~\ref{eq:momentum_flux}
can be expressed as
\begin{align}
  \label{eq:integral1}
  \vec{p}_\nu(\theta_\mathrm{n})& =
  \frac{I_{\nu}}{c}\bar{\vec{n}}_{\mathrm{d}\mathcal{A}} \left(
  \int_0^{2\pi} \mathrm{d}\phi \int_0^{\pi/2} \mathrm{d}\theta
  \cos^2{\theta} \sin{\theta} \, \right.\\ & \quad\label{eq:integral2}
  \left.-2\int_{\pi/2-\theta_{\mathrm{n}}}^{\pi/2}
  \mathrm{d}\theta\cos^{2}{\theta} \sin{\theta}
  \int_{\phi_{\mathrm{lim},1}}^{\phi_{\mathrm{lim},2}}
  \,\mathrm{d}\phi \right) \\ &
  =\frac{I_{\nu}}{c}\bar{\vec{n}}_{\mathrm{d}\mathcal{A}}\left(\frac{2\pi}
  {3}-2\int_{\pi/2-\theta_{\mathrm{n}}}^{\pi/2}\mathrm{d}\theta
  \cos^{2}{\theta} \sin{\theta}
  \int_{\phi_{\mathrm{lim},1}}^{\phi_{\mathrm{lim},2}} \,
  \mathrm{d}\phi \right) ,
\end{align}
 where the first integral (Eq.~\ref{eq:integral1}) corresponds to the
 hemisphere over the surface element $\mathrm{d}\mathcal{A}$ in
 Fig.~\ref{fig:pressure_geometry} and the second integral
 (Eq.~\ref{eq:integral2}) corresponds to the two green parts. The
 first of the latter is subtracted because it lies below the hotspot
 surface and does not contribute, the second one has a negative
 contribution to the momentum flux because it is below the surface
 element d$\mathcal{A}$. These two regions are identical by symmetry
 and allow us to use the same limits of integration in $\phi$ and
 $\theta$.

Our estimate of the opening angle $\theta_{\mathrm{flow}}$ of the
outflow is based on the integral of the momentum flux $\vec{p}_\nu$
over $\theta_\mathrm{n}$. We define it as the angle that corresponds
to a fraction of 0.68 (arbitrarily set; standard $1\sigma$ as for a
Gaussian distribution) of the total integral, i.\,e.,
\begin{equation}
  \int_{0}^{\theta_\mathrm{flow}/2}\vec{p}_\nu(\theta_{\mathrm{n}})\,\mathrm{d}\theta_{\mathrm{n}}=
  0.68\int_0^{\pi/2}\vec{p}_\nu(\theta_\mathrm{n})\,\mathrm{d}\theta_\mathrm{n}.
\end{equation}
This gives an opening angle of $\theta_{\mathrm{flow}}
\sim90^{\circ}$. The `effective' solid angle in which the disc-like
outflow spreads is given by
\begin{equation}
\label{eq:sigma}
  4\pi f_\mathrm{c} = \int_{0}^{2\pi}\mathrm{d}\phi
  \int_{\pi/2-\theta_{\mathrm{flow}}/2}^{\pi/2+\theta_{\mathrm{flow}}/2}\mathrm{d}\theta
  \sin\theta \ = 4\pi\sin(\theta_\mathrm{flow}/2) \sim 2.84\pi\,
  \mathrm{sr} ,
\end{equation}
where $f_\mathrm{c}$ is the covering factor that is defined as the
fraction of the spherical domain that is covered by gas
(Sect.~\ref{sec:rho}). This is different from the geometry adopted in
Paper~I, where the outflow was arbitrarily assumed to be confined in a
quarter of a sphere (solid angle equal to $\pi$).

\section{Calculation of the effective flux}
\label{sec:flux}
\begin{figure}[t]
  \centering
  \includegraphics[width=0.4\textwidth]{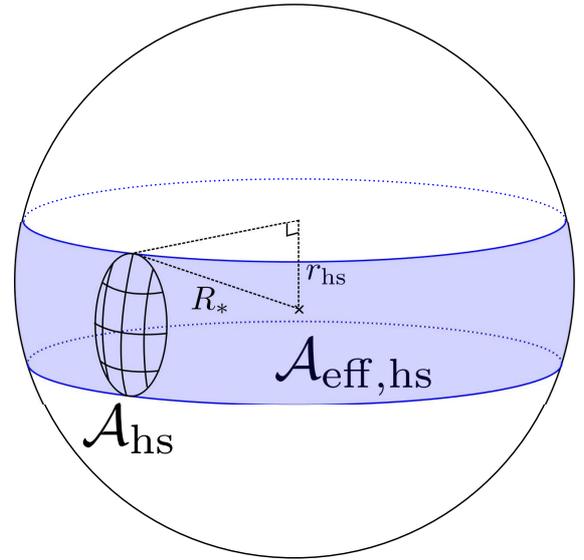}
  \caption{Representation of the effective surface of the hotspot over
    a full orbital period. See text for description.}
  \label{fig:geometry}
\end{figure}
In our simulations, the effective radiative flux $F_\mathrm{eff}$ is
given by
\begin{equation}
  F_\mathrm{eff} = F_\mathrm{g} +
  \frac{\mathcal{A}_\mathrm{hs}}{\mathcal{A}_\mathrm{eff,hs}}
  F_\mathrm{hs},
\end{equation}
where $F_{\mathrm{g}}=L_*/\mathcal{A}_*$ and
$F_{\mathrm{hs}}=L_\mathrm{hs}/\mathcal{A}_\mathrm{hs}$ are the gainer
star and hotspot fluxes, respectively, $\mathcal{A}_{*}$ is the
surface area of the star, and $\mathcal{A}_{\mathrm{hs}}$ is the
surface area of the hotspot, i.\,e.\ the surface of a calotte with the
radius of the base equivalent to the radius of the impacting stream,
\begin{equation}
  \label{ap:aire}
  \mathcal{A}_\mathrm{hs}=2\pi
  R_{*}\left(R_{*}-\sqrt{R_{*}^{2}-r_{\mathrm{hs}}^{2}}\right),
\end{equation}
where $R_{*}$ and $r_{\mathrm{hs}}$ are the star and hotspot
radius. $\mathcal{A}_\mathrm{eff,hs}$ is the area of the strip at the
equator of the star of height 2$r_{\mathrm{hs}}$
(Fig.~\ref{fig:geometry}),
\begin{equation}
  \mathcal{A}_{\mathrm{eff,hs}} = 4 \pi R_{*}^{2} - 2 \times 2 \pi
  R_{*} (R_{*} - r_{\mathrm{hs}}) ,
\end{equation}
over which the hotspot flux is averaged.
Therefore, the effective flux reads
\begin{equation}
  F_{\mathrm{eff}} = F_{\mathrm{g}} + \frac{R_{*}}{2 r_{\mathrm{hs}}} \left[1
  - \sqrt{1 - \left(\frac{r_{\mathrm{hs}}}{R_{*}} \right)^{2}} \right]
  F_{\mathrm{hs}} .
\end{equation}
The net flux from the hotspot is thus reduced by the factor
\begin{equation}
  \gamma = \frac{R_{*}}{2 r_{\mathrm{hs}}} \left[1 - \sqrt{1 -
  \left(\frac{r_{\mathrm{hs}}}{R_{*}} \right)^{2}} \right].
\end{equation}
In our simulations, the source of
radiation emits isotropically, so that the total effective luminosity
will be
\begin{equation}
    L_\mathrm{eff}=L_\mathrm{g}+\mathcal{A}_*F_\mathrm{eff,hs}
            =L_\mathrm{g}+\frac{R_*}{r_\mathrm{hs}}L_\mathrm{hs},
\end{equation}
where $F_\mathrm{eff,hs }=
(\mathcal{A}_\mathrm{hs}/\mathcal{A}_\mathrm{eff,hs}) F_\mathrm{hs}$.

\section{UV continuum and line variability in W~Ser}
\label{ap:WSer_UV}

\begin{figure}[t]
  \centering\includegraphics[width=0.49\textwidth]{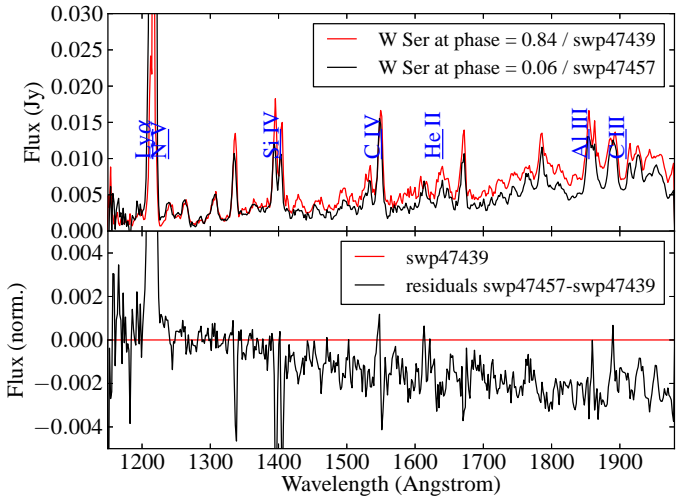}
  \caption{\textbf{Top:} Continuum in the UV band for W~Ser observed with
    IUE. Black: SWP47439 observation made on 1993-04-07 (20:02:16) at
    phase $\Phi = 0.84$; red: SWP47457 observation made on 1993-04-10
    (21:40:17) at phase $\Phi = 0.06$. \textbf{Bottom:} Difference of the
    two spectra.}
  \label{fig:WSer_UV}
\end{figure}
\begin{figure}[t]
  \centering\includegraphics[width=0.49\textwidth]{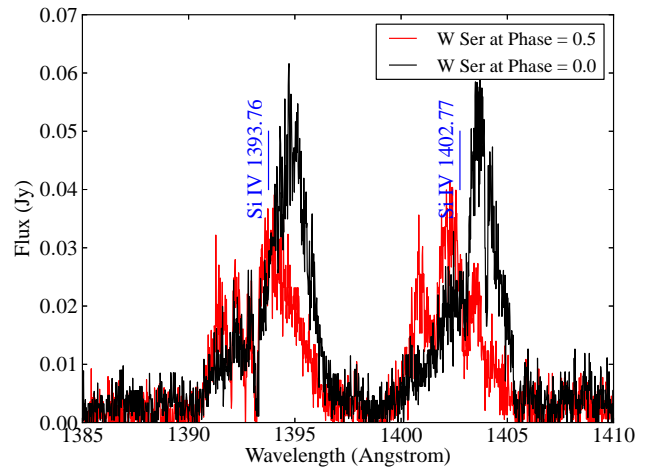}
  \caption{HST/GHRS spectra of the [\ion{Si}{iv}] 1393.76 and
    1402.77~{\AA} lines of W~Ser at two distinct phases (black:
    z0lu5108t observation made on 1991-07-12 (05:32:25) at phase
    $\Phi= 0.00$, primary eclipse; red: z0lu0308t observation made on
    1991-07-19 (03:29:13) at phase $\Phi = 0.5$, secondary eclipse).}
  \label{fig:W_SerSi}
\end{figure}

W~Ser systems are known to show a strong line variability in the UV
(especially for highly ionised species; see the review by
\citealt{1982NASCP2338..526P}). In this appendix, we investigate, for
the prototypical system W~Ser, the observed UV spectra in the search
for variations in the continuum as presented in
Sect.~\ref{sec:output_spectra}. We also demonstrate the variability of
[\ion{Si}{iv}] lines for this object.

Figure~\ref{fig:WSer_UV} shows low-dispersion, large aperture spectra
of W~Ser in the UV as observed by the International Ultraviolet
Explorer (IUE) at two distinct phases (these phases are estimated
thanks to the improved ephemeris of
\citealt{2005ApJ...632..576P}). The two spectra have been taken only 3
days apart limiting the effect of intrinsic variability.  The red
spectrum is taken between the secondary and the primary eclipse at
phase 0.8 and the black one at phase 0.1 during the primary
eclipse. We note that at phase 0.1, the absorption in the longer
wavelength region of the spectra ($\lambda > 1400$ \AA) and the
Lyman~$\alpha$ emission are stronger than during the secondary
eclipse, possibly indicative of a hot ionised region.

Figure~\ref{fig:W_SerSi} presents the profiles of the [\ion{Si}{iv}]
1393.76 and 1402.77~{\AA} lines as observed with HST/GHRS for W~Ser at
two different phases: primary eclipse ($\Phi=0.0$), and secondary
eclipse ($\Phi=0.5$). The same ephemeris has been used. We clearly see
an excess of flux in the blue wing at phase $\Phi=0.5$. The total
extent of the wings of both lines reaches 600~km\,s$^{-1}$, which is
comparable with the escape velocity at the surface of the star
(643~km\,s$^{-1}$ considering a main sequence gainer star of
1.51~$M_\odot$ according to \citealt{2004A&A...417..263B}, and
following a mass-radius relation for main-sequence stars of the form
$R\propto M^{0.8}$) and with the outflow terminal velocity in our
model (850~km\,s$^{-1}$). In their attempt to model this line
variability, \cite{1995ApJ...447..401W} found two components, a broad
one likely due to the boundary layer of an accretion disc and a narrow
one possibly due to the formation of a hotspot. Further investigation,
especially at different phases, may shed new light on the origin of
the [\ion{Si}{iv}] emission.

\section{Chemical composition of the outflow}

Table~\ref{tab:composition} presents the chemical composition of the
donor star atmosphere that will be transferred by RLOF onto the
accretor's surface and eventually ejected by the hotspot.
\begin{table*}
  \caption{Chemical composition of the outflowing material for the
    three models computed.}  \centering
  \begin{tabular}{ccccc}
    \hline
    \hline\\
    Element ($i$) & $M_{i}/M_{\mathrm{tot}}$ & $M_{i}/M_{\mathrm{tot}}$ & $M_{i}/M_{\mathrm{tot}}$ & $M_{i}/M_\mathrm{tot}$ \\
    & (Model A) & (Model B ) & (Model C) & (Solar ref.) \\ \\
    \hline\\
    H    & $6.9941 \times 10^{-1}$ & $6.9902 \times 10^{-1}$ & $6.9810 \times 10^{-1}$ & $7.3732\times10^{-1}$ \\
    He   & $2.8059 \times 10^{-1}$ & $2.8074 \times 10^{-1}$ & $2.8134 \times 10^{-1}$ & $2.4917\times10^{-1}$ \\
    Li   & $6.7290 \times 10^{-18}$& $3.1061 \times 10^{-18}$& $1.0826 \times 10^{-18}$ & $1.0358\times10^{-8}$ \\
    C    & $3.4675 \times 10^{-3}$ & $2.2462 \times 10^{-3}$ & $1.0944 \times 10^{-4}$ & $2.1526\times10^{-3}$ \\
    N    & $1.0632 \times 10^{-3}$ & $2.5270 \times 10^{-3}$ & $4.9937 \times 10^{-3}$ & $8.7194\times10^{-4}$ \\
    O    & $9.6519 \times 10^{-3}$ & $9.6486 \times 10^{-3}$ & $9.6333 \times 10^{-3}$ & $5.7348\times10^{-3}$ \\
    F    & $5.6122 \times 10^{-7}$ & $5.6411 \times 10^{-7}$ & $5.6984 \times 10^{-7}$ & $4.1970\times10^{-7}$ \\
    Ne   & $1.9697 \times 10^{-3}$ & $1.9696 \times 10^{-3}$ & $1.9575 \times 10^{-3}$ & $1.4762\times10^{-3}$ \\
    Na   & $4.0004 \times 10^{-5}$ & $4.0130 \times 10^{-5}$ & $5.2729 \times 10^{-5}$ & $3.5989\times10^{-5}$ \\
    Mg   & $7.5221 \times 10^{-4}$ & $7.5221 \times 10^{-4}$ & $7.5220 \times 10^{-4}$ & $6.1694\times10^{-4}$ \\
    Al   & $6.4825 \times 10^{-5}$ & $6.4825 \times 10^{-5}$ & $6.4825 \times 10^{-5}$ & $5.8225\times10^{-5}$ \\
    Si   & $8.1136 \times 10^{-4}$ & $8.1136 \times 10^{-4}$ & $8.1136 \times 10^{-4}$ & $7.1290\times10^{-4}$ \\
    P    & $7.1102 \times 10^{-6}$ & $7.1102 \times 10^{-6}$ & $7.1102 \times 10^{-6}$ & $7.2504\times10^{-6}$ \\
    S    & $4.3015 \times 10^{-4}$ & $4.3015 \times 10^{-4}$ & $4.3015 \times 10^{-4}$ & $4.3159\times10^{-4}$ \\
    Cl   & $2.3983 \times 10^{-6}$ & $2.3983 \times 10^{-6}$ & $2.3983 \times 10^{-6}$ & $4.9535\times10^{-6}$ \\
    heavy& $1.7395 \times 10^{-3}$ & $1.7395 \times 10^{-3}$ & $1.7395 \times 10^{-3}$ &   \\ \\
    \hline
  \end{tabular}
  \label{tab:composition}
  \tablefoot{Contributions from different isotopes (when available)
    are added up. For the \textsc{Cloudy} simulations we use the
    thirty lightest elements. For those not appearing in the table we
    adopt solar abundances.}  \tablebib{Solar abundances:
    \citet{1998SSRv...85..161G}, except N, Ne, Mg, Si, and Fe
    \citep{2001AIPC..598...23H}, O \citep{2001ApJ...556L..63A}, and C
    \citep{2002ApJ...573L.137A}.}
\end{table*}
\end{appendix}

\bibliographystyle{aa}
\bibliography{papier2}
\end{document}